\pdfoutput=1
\documentclass[11pt,a4paper]{article}

\usepackage[utf8]{inputenc}
\usepackage[english]{babel}
\usepackage[T1]{fontenc}
\usepackage[margin=3.2cm]{geometry}

\usepackage{amsmath,amssymb,amsfonts,amsthm,mathrsfs,bm}
\usepackage{dsfont}
\usepackage{graphicx}
\usepackage{booktabs}
\usepackage{float}
\usepackage{caption}
\usepackage{subfig}
\usepackage[square,numbers,sort&compress]{natbib}
\usepackage[hidelinks]{hyperref}
\usepackage{titlesec}
\usepackage{color}

\usepackage{etoolbox}
\apptocmd{\thebibliography}{\raggedright}{}{}

\usepackage{algorithm}
\usepackage{algpseudocode}
\algrenewcommand\algorithmicrequire{\textbf{Input:}}
\algrenewcommand\algorithmicensure{\textbf{Output:}}

\theoremstyle{definition}

\newtheorem{proposition}{Proposition}

\providecommand{\Description}[1]{}
\title{Confidence intervals for empirical convergence rates of
       randomised quasi-Monte Carlo, with applications to option pricing}
\author{Giacomo Case \\ \normalsize EPFL, Lausanne, Switzerland}
\date{}

\begin{document}
\maketitle

\begin{abstract}
Empirical comparisons of quasi-Monte Carlo rules often report fitted
convergence exponents without intervals. We estimate the root mean square error
from independent randomisations, fit its log slope over a declared sample-size
window, and resample whole randomisations to construct an interval. For a fixed
window, we establish asymptotic validity under finite fourth moments and a
positive limiting variance. We measure coverage of four interval constructions
on integrands with known exponents. With Gaussian errors, all four are
compatible with nominal coverage at 128 and 512 randomisations, and
jackknife-$t$ already at 32. All four undercover for the two higher-kurtosis
families even at 512 randomisations. For differences between slopes,
including null and small effects, preserving the dependence between paired
randomisations can greatly narrow intervals, but it does not ensure nominal
coverage. In option pricing, a steeper slope and a smaller error at a
fixed budget can rank rules differently. Preintegration
lowers a digital Asian option's exponent by about 0.5 in every tested window.
For a barrier option it reduces the error six- to ninefold while changing the
exponent by less than four hundredths. An arithmetic basket's gain grows with
sample size. A geometric basket remains a ridge function as assets are added,
limiting its use as a dimension benchmark. For multi-asset Asian options,
different bases within a degenerate PCA eigenspace can change the measured
exponent, in a direction that depends on the payoff.
\end{abstract}
\section{Introduction}
\label{sec:introduction}

Pricing a path-dependent or multi-asset contract by simulation means computing
an integral whose dimension is set by the contract. In the discretely monitored
Black--Scholes models of this paper there is one Gaussian coordinate for each
asset and each monitoring date. For an integrand with finite variance, the root
mean square error of plain Monte Carlo is of order $n^{-1/2}$ in every
dimension. Quasi-Monte Carlo replaces the random sample by a low-discrepancy
point set and often converges faster \citep{DickKuoSloan2013,Paul}. How much
faster depends on the integrand, and applied studies often report the rate as an
empirical convergence exponent, the slope of the log error against the log
sample size \citep{Kucherenko2011effective,Bianchetti2015pricing}.

The fitted exponent depends on the randomisations and the range of sample sizes
used for the fit. Comparing two methods requires an interval on the difference
of their exponents that accounts for dependence between the runs. For
randomised quasi-Monte Carlo estimates of an integral, confidence intervals
have been studied in detail
\citep{LEcuyerNakayamaOwenTuffin2023,NakayamaTuffin2024,Owen2026error}. For a
fitted exponent we have not found a study of interval coverage.

This paper gives a protocol for estimating and comparing empirical convergence
exponents of randomised quasi-Monte Carlo rules and applies it to option
pricing. We provide:
\begin{itemize}
\item A protocol in which the error is estimated from $R$ independent
  randomisations, with nested point sets where the rule allows it, the exponent
  is fitted over a declared window of sample sizes, and its interval resamples
  whole randomisations (Section~\ref{sec:protocol}).
\item A definition of the finite-window slope that the intervals estimate,
  with conditions under which they are valid as the number of randomisations
  grows (Section~\ref{sec:target}), and a measurement of the coverage of four
  interval constructions on integrands with a known exponent and replicate
  errors that range from Gaussian to heavy-tailed
  (Section~\ref{sec:known-answer}).
\item Intervals on the difference between the exponents of two rules, paired
  when the rules share seeds, with their coverage measured on integrands whose
  difference is known (Section~\ref{sec:difference-calibration}), together with
  fit-window sensitivity and comparisons at estimated equal cost
  (Section~\ref{sec:windows}).
\item Applications to option pricing: a reassessment of a standard
  basket benchmark (Sections~\ref{sec:corrections} and~\ref{sec:ridge}), an
  example in which low effective dimension does not give a fast rate
  (Section~\ref{sec:decoupling}), the effect of preintegration on the exponent
  and on the size of the error (Sections~\ref{sec:preint}
  and~\ref{sec:windows}), and the role of the eigenbasis for multi-asset
  contracts (Section~\ref{sec:multiasset}).
\end{itemize}

\subsection{Related work}

\emph{Error estimation for randomised quasi-Monte Carlo.} Independent
randomisations of a quasi-Monte Carlo rule give an unbiased estimate of the
integral together with a sample from which its error can be estimated
\citep{Owen2023practical,Owen2026error}. L'Ecuyer, Nakayama, Owen and
Tuffin~\citep{LEcuyerNakayamaOwenTuffin2023} compare confidence intervals for the
mean of such estimates and find Student intervals more reliable than the
bootstrap intervals they tested, including the bootstrap-$t$ interval. Nakayama
and Tuffin~\citep{NakayamaTuffin2024} give sufficient conditions for central
limit theorems and valid confidence intervals. These works concern the value of
an integral. Here the target is a slope fitted to estimated errors at several
sample sizes, whose dependence must be retained when constructing its interval.

\emph{Empirical convergence rates in applications.} Early applications of
quasi-Monte Carlo to derivative pricing found it effective on integrals of high
nominal dimension \citep{Corwin}. Applied studies commonly report convergence
exponents read from fitted trend lines and compare them across methods and path
constructions
\citep{Kucherenko2011effective,Bianchetti2015pricing,ScoleriBianchettiKucherenko2021}.
We assess the uncertainty of these comparisons and their sensitivity to the
fit window.

\emph{Effective dimension and smoothness.} The success of quasi-Monte Carlo in
high nominal dimension is usually explained by a low effective dimension.
Caflisch, Morokoff and Owen introduced the notion and lowered it with the
Brownian bridge \citep{CaflischMorokoffOwen1997}, and Wang and
Fang~\citep{WangFang2003} studied its relation to quasi-Monte Carlo integration.
Principal components and the linear transformation of Imai and
Tan~\citep{ImaiTan2006} pursue the same reduction by other rotations, and
tractability in weighted spaces supplies the corresponding theory
\citep{SloanWozniakowski1998,DickKuoSloan2013}. Owen~\citep{Owen2003dimension}
showed that low effective dimension alone does not make quasi-Monte Carlo much
better than Monte Carlo. For discontinuous payoffs, Wang and
Tan~\citep{WangTan2013} found that the structure of the discontinuities can
matter more than the effective dimension, and He and
Wang~\citep{HeWang2015discontinuous} give convergence rates of randomised
quasi-Monte Carlo for discontinuous functions. We also examine how the choice
of eigenbasis inside a degenerate eigenspace changes the measured exponent of
a multi-asset contract.

\emph{Preintegration and conditioning.} Integrating out a suitably chosen
variable can smooth a discontinuous integrand. He~\citep{He2019rate} shows that
conditional randomised quasi-Monte Carlo attains an error of
$O(n^{-1+\epsilon})$ for some typical option-pricing problems, and Wang and
Tan~\citep{WangTan2013} rotate the coordinates so that a discontinuity becomes
parallel to the axes. Methods that act on dimension and smoothness together
include the linear transformation with a realignment of the discontinuity of
Imai and Tan~\citep{ImaiTan2014}, the conditional method with dimension
reduction of Xiao and Wang~\citep{XiaoWang2018}, and the integrated method of He
and Wang~\citep{HeWang2017}. Smoothing by preintegration
\citep{GriewankKuoLeoveySloan2018} is the form used here. Gilbert, Kuo and
Sloan~\citep{GilbertKuoSloan2021} study its monotonicity hypothesis and what
happens without it, and Liu~\citep{Liu2022} chooses the conditioning direction
through constrained active subspaces. For barrier options, Achtsis, Cools and
Nuyens~\citep{AchtsisCoolsNuyens2012} develop conditional sampling under the
Heston model. We condition on the leading principal component and measure what
this changes in the fitted exponent and in the size of the error, over several
fit windows and at an estimated equal cost.

\subsection{Notation}
\label{sec:intro-counters}

We use $n$ for the sample size, $d$ for the number
of assets in a basket, $m$ for the number of monitoring dates of a path-dependent
contract, $R$ for the number of independent randomisations of a rule, and $C$
for the number of contracts in a family. In the general definitions of
Sections~\ref{sec:sec_and_subsec}, \ref{sec:discrepancy} and~\ref{sec:anova},
$d$ also denotes the dimension of the integration domain. For the pricing
integrands that dimension is $d$, $m$ or $d\,m$, as stated where they appear.
Other symbols are introduced where they are first used and collected in
Appendix~\ref{app:notation}.

\section{Monte Carlo and quasi-Monte Carlo integration}
\label{sec:sec_and_subsec}

We want $\alpha=\mathbb{E}[f(\mathbf{U})]=\int_{[0,1)^d}f(\mathbf{x})\,d\mathbf{x}$
for $\mathbf{U}$ uniform on $[0,1)^d$. Monte Carlo draws $n$ independent uniform
points $\mathbf{U}_1,\dots,\mathbf{U}_n$ and uses
\begin{equation}
  \hat{\alpha}_n=\frac{1}{n}\sum_{j=1}^n f(\mathbf{U}_j),
\end{equation}
whose error has standard deviation $\sigma(f)/\sqrt{n}$ for every $f$ with
finite variance and every $d$. Quasi-Monte Carlo takes the same average over
deterministic points $\mathbf{x}_1,\dots,\mathbf{x}_n$ chosen to fill the cube
evenly. Its error is a number rather than a random variable, and the
Koksma--Hlawka inequality bounds it by the product of a term that depends only
on the points and a term that depends only on the integrand,
\begin{equation}
  \Bigl|\alpha-\tfrac{1}{n}\textstyle\sum_{i=1}^{n}f(\mathbf{x}_i)\Bigr|
  \le \mathcal{D}^*(\mathbf{x}_1,\dots,\mathbf{x}_n)\,V(f),
\end{equation}
where $\mathcal{D}^*$ is the star discrepancy of the points, defined in
Section~\ref{sec:discrepancy}, and $V(f)$ is the variation of $f$ in the sense
of Hardy and Krause. The inequality gives a useful bound only when $V(f)$ is
finite. After the inverse normal transformation the unconditioned call payoffs
are unbounded, and the digital and barrier payoffs jump across surfaces that are
not parallel to the coordinate axes. Either property makes the variation
infinite, although a jump alone does not when it lies along the axes. The
inequality therefore gives no bound for these integrands, which is one reason
the paper measures rates instead of deriving them.

\section{The test contracts}
\label{sec:n_simu}

\subsection{Geometric basket call option}

Options on the geometric average of lognormal asset prices are a standard test
case for quasi-Monte Carlo, because their price has a closed form \citep{Paul}.
The payoff is $(\Bar{S}_{\mathrm{geo}}-K)^+$, with
\[
  \Bar{S}_{\mathrm{geo}}=\Bigl(\prod_{i=1}^d S_i(T)\Bigr)^{1/d},
\]
where $S_1,\dots,S_d$ are the asset prices, $T$ is the maturity and $K$ the
strike. For this single-date contract $d$ is both the number of assets and the
dimension of the integral. Under the risk-neutral measure each asset follows
\[
  \frac{dS_i(t)}{S_i(t)}=r\,dt+\sigma_i\,dW_i(t),
\]
with risk-free rate $r$, volatility $\sigma_i$ and standard Brownian motions
with $d\langle W_i,W_j\rangle(t)=\rho_{ij}\,dt$, where $\rho_{ii}=1$. The
geometric average is again lognormal,
\[
  \Bar{S}_{\mathrm{geo}}=\Bar{S}_{\mathrm{geo}}(0)\,
  \exp\Bigl\{\Bigl[r-\frac{1}{2d}\sum_{i=1}^d\sigma_i^2\Bigr]T
  +\frac{1}{d}\sum_{i=1}^d\sigma_iW_i(T)\Bigr\},
  \qquad
  \Bar{S}_{\mathrm{geo}}(0)=\Bigl(\prod_{i=1}^d S_i(0)\Bigr)^{1/d},
\]
with
$\sum_{i=1}^d\sigma_iW_i(T)\sim\mathcal{N}\bigl(0,\,T\sum_{i=1}^d\sum_{j=1}^d\rho_{ij}\sigma_i\sigma_j\bigr)$.
Its price therefore has a closed form, given in Appendix~\ref{app:closedform}.
We use it to check the implementation and to measure errors against the exact
price in Section~\ref{sec:corrections} and Appendix~\ref{app:faure}.

\subsection{Asian call option}
\label{sec:asian-contract}

An Asian call option on a single lognormal asset pays
$(\Bar{S}_{\mathrm{ari}}-K)^+$ at $T$, where
\[
  \Bar{S}_{\mathrm{ari}}=\frac{1}{m}\sum_{i=1}^m S(t_i)
\]
is the average over fixed monitoring dates $0<t_1<\dots<t_m\le T$. Its value is
\[
  PV_0=e^{-rT}\int_{[0,1)^m}\Bigl(\frac{1}{m}\sum_{i=1}^m S(0)\,e^{h_i(\mathbf{x})}-K\Bigr)^+\,d\mathbf{x},
\]
where $t_0=0$ and
\[
  h_i(\mathbf{x})=\Bigl(r-\frac{\sigma^2}{2}\Bigr)t_i
  +\sigma\sum_{\ell=1}^i\sqrt{t_\ell-t_{\ell-1}}\,\Phi^{-1}(x_\ell).
\]
This form uses the incremental path construction, in which coordinate $\ell$
drives the Brownian increment over $(t_{\ell-1},t_\ell]$. The
arithmetic average is a sum of dependent lognormal variables, and the option has
no known closed-form price.

\subsection{Figure of merit}
\label{sec:figure-of-merit}

For a family of $C$ contracts with exact values $\chi_1,\dots,\chi_C$ and
estimates $\hat\chi_1(n),\dots,\hat\chi_C(n)$ from $n$ points, the root mean
square error over the family is
\[
  \mathrm{RMSE}(n)=\Bigl[\frac{1}{C}\sum_{i=1}^C\bigl(\hat\chi_i(n)-\chi_i\bigr)^2\Bigr]^{1/2}.
\]
For a deterministic point set this is one number per sample size. For a random
rule each $\hat\chi_i(n)$ is a random variable, and the squared error is
replaced by its expectation over the randomisation before the square root is
taken. For unbiased rules this quantity can be estimated
from the spread of independent randomisations, without reference prices
(Section~\ref{sec:protocol}).

\section{Discrepancy}
\label{sec:discrepancy}

\subsection{Definitions and bounds}

For a family $\mathscr{A}$ of boxes in $[0,1)^d$, the discrepancy of a point set
$P=\{\mathbf{x}_1,\dots,\mathbf{x}_n\}$ with respect to $\mathscr{A}$ is the
largest difference between the fraction of points in a box and the volume of
the box \citep{Niederreiter1,Paul},
\[
  \sup_{A\in\mathscr{A}}\Bigl|\frac{\#\{i:\mathbf{x}_i\in A\}}{n}-\mathrm{vol}(A)\Bigr|.
\]
All boxes $\prod_{j=1}^d[u_j,v_j)$ give the extreme discrepancy
$\mathcal{D}(P)$, and the boxes anchored at the origin, $\prod_{j=1}^d[0,v_j)$,
give the star discrepancy $\mathcal{D}^*(P)$. The two agree up to a factor that
depends only on the dimension,
$\mathcal{D}^*(P)\le\mathcal{D}(P)\le 2^d\mathcal{D}^*(P)$ \citep{Niederreiter1}.
The star discrepancy is the one in the Koksma--Hlawka inequality of
Section~\ref{sec:sec_and_subsec}. For a sequence we write $\mathcal{D}^*_n$ for
the star discrepancy of its first $n$ points.

The discrepancy of $n$ points cannot decay arbitrarily fast. For point sets in
$d\ge 2$ dimensions it is conjectured that
$\mathcal{D}^*(P)\ge c_d(\log n)^{d-1}/n$, and for sequences that
$\mathcal{D}^*_n\ge c_d'(\log n)^{d}/n$ for infinitely many $n$, with constants
that depend only on $d$ \citep{Niederreiter}. The first bound is a theorem for
$d=2$ and the second for $d=1$. In higher dimension Roth's bound, with the
exponent $(d-1)/2$ in place of $d-1$, holds for every point set
\citep{Niederreiter1}, and Bilyk, Lacey and
Vagharshakyan~\citep{BilykLaceyVagharshakyan2008} improve that exponent by a
small amount that depends on $d$. A sequence with
$\mathcal{D}^*_n=O((\log n)^d/n)$ is called a low-discrepancy sequence. Since
$(\log n)^d/n=O(n^{-1+\epsilon})$ for every $\epsilon>0$, the Koksma--Hlawka
inequality gives such a sequence an error of order $o(n^{-1/2})$ in every fixed
dimension, for integrands of bounded variation. The constant and the factor
$(\log n)^d$ can make this comparison uninformative at practical sample sizes.

\subsection{Nets and sequences}

Most constructions used here are digital nets and sequences, whose uniformity is
defined through elementary intervals \citep{Niederreiter1,DickKuoSloan2013}. In a
base $b\ge 2$, an elementary interval is a box
\[
  E=\prod_{j=1}^d\Bigl[\frac{c_j}{b^{k_j}},\frac{c_j+1}{b^{k_j}}\Bigr),
  \qquad k_j\ge 0,\quad 0\le c_j<b^{k_j},
\]
of volume $b^{-(k_1+\dots+k_d)}$. For integers $0\le t\le k$, a set of $b^k$
points is a $(t,k,d)$-net in base $b$ if every elementary interval of volume
$b^{t-k}$ contains exactly $b^t$ of its points. The net then integrates the
indicator of every such interval without error, and a smaller $t$ means exact
integration on finer intervals. A sequence is a $(t,d)$-sequence in base $b$ if,
for every $k\ge t$ and every $j\ge 0$, the $b^k$ consecutive points with indices
$jb^k,\dots,(j+1)b^k-1$ form a $(t,k,d)$-net. In the literature these are
usually written as $(t,m,s)$-nets and $(t,s)$-sequences, and we keep $d$ for the
dimension and $m$ for the number of monitoring dates. The first $b^k$ points
are the block with $j=0$. This is why sample sizes that are powers of the base
suit such sequences, and why a block that does not start at a multiple of $b^k$
need not be a net.

For fixed $t$, $b$ and $d$ these nets and sequences have low discrepancy. A
$(t,k,d)$-net in base $b$ with $n=b^k$ points has
$\mathcal{D}^*(P)=O(b^t(\log n)^{d-1}/n)$, and a $(t,d)$-sequence has
$\mathcal{D}^*_n=O(b^t(\log n)^d/n)$, with constants that depend on $b$ and $d$
\citep{Niederreiter1}. The factor $b^t$ penalises a large quality parameter $t$,
and the base affects both the constant and the sample sizes at which the net
property holds.

\subsection{Discrepancy in high dimension}
\label{sec:high-dim}

These bounds say little about the Asian contracts. The factor $(\log n)^d/n$
increases with $n$ up to $n=e^d$, and for $d=121$ it exceeds $10^{130}$ at
$n=2^{20}$. The unconditioned pricing integrands also have infinite variation
(Section~\ref{sec:sec_and_subsec}). The bounds therefore do not explain why
quasi-Monte Carlo performs well on many such integrands \citep{Corwin}. The usual
explanation is that the integrands depend mostly on a few coordinates, or on
interactions of low order. Weighted function spaces formalise this idea. When
the importance of the coordinates decreases fast enough, the worst-case error
can be bounded independently of $d$ \citep{SloanWozniakowski1998,DickKuoSloan2013}.
The effective dimension measured in Section~\ref{sec:decoupling} is an empirical
counterpart of that idea.

\subsection{Van der Corput sequences}

Every non-negative integer $\omega$ has a unique expansion in a base $b\ge 2$,
\begin{equation}
  \label{eq:baseb}
  \omega=\sum_{k\ge 0}a_k(\omega)\,b^k,\qquad a_k(\omega)\in\{0,1,\dots,b-1\}.
\end{equation}
The radical inverse reflects the digits about the radix point,
\begin{equation}
  \label{eq:radinv}
  \psi_b(\omega)=\sum_{k\ge 0}\frac{a_k(\omega)}{b^{k+1}},
\end{equation}
and the van der Corput sequence in base $b$ is $\psi_b(0),\psi_b(1),\dots$. It
is a $(0,1)$-sequence in base $b$ \citep{Niederreiter1}, so its first $b^k$
points put one point in each of the $b^k$ intervals of length $b^{-k}$. At a
fixed resolution of $k$ digits a larger base therefore needs more points before
every interval is filled.

\subsection{Low-discrepancy sequences}

\subsubsection{Halton}

Let $b_1,\dots,b_d$ be the first $d$ primes. The Halton sequence \citep{Halton}
has the points
$\mathbf{x}_\omega=(\psi_{b_1}(\omega),\dots,\psi_{b_d}(\omega))$. Its
coordinates use different bases, so it is not a net in any single base, but it
is a low-discrepancy sequence. The constant in its discrepancy bound grows
superexponentially with $d$ \citep{Niederreiter1}, and for moderate $n$ the
two-dimensional projections of coordinates with large bases show strong
correlations, which led Kocis and Whiten~\citep{Kocis} to propose a leaped
variant. We use the randomisation of
Owen~\citep{Owen2017Halton}, which applies an independent random permutation of
the digits to each digit position of each coordinate. Conclusions about
deterministic Halton points need not apply to this randomised construction.

\subsubsection{Faure}
\label{sec:faure-base}

The Faure sequence \citep{Faure} uses a single prime base $b\ge d$ for every
coordinate. With the base-$b$ digits $a_\kappa(\omega)$ of \eqref{eq:baseb},
coordinate $i$ of point $\omega$ is
\begin{equation}
  y(i,\omega)=\sum_{j\ge 1}\frac{y_j(i,\omega)}{b^j},\qquad
  y_j(i,\omega)=\sum_{\kappa\ge j-1}\binom{\kappa}{j-1}(i-1)^{\kappa-j+1}a_\kappa(\omega)\bmod b,
\end{equation}
with $0^0=1$, so that the first coordinate is the van der Corput sequence in
base $b$. The Faure sequence is a $(0,d)$-sequence in base $b$
\citep{Faure,Niederreiter1}, so its quality parameter takes the smallest possible
value. Its base, however, grows with the dimension. A block of $b^2$ points has
one point in each cell of side $b^{-1}$ in every pair of coordinates, and a
sample with fewer than $b^2$ points cannot have this property, although it can
still be spread evenly in other respects. We use the smallest prime $b\ge d$.

\subsubsection{Sobol'}

Sobol' points \citep{Sobol} are built in base $2$. Each coordinate has its own
generator matrix $\mathbf{G}$ over $\mathbb{F}_2$, upper triangular with ones on
the diagonal. In binary index order, writing $\mathbf{a}(\omega)$ for the
binary digits of $\omega$,
the binary digits of the coordinate's $\omega$-th value are
$\mathbf{y}(\omega)=\mathbf{G}\mathbf{a}(\omega)\bmod 2$, and the value itself is
$\sum_{j\ge 1}y_j(\omega)2^{-j}$. The columns of $\mathbf{G}$ are the binary
expansions of the direction numbers $\nu_1,\nu_2,\dots$ of the coordinate, which
are derived from primitive polynomials over $\mathbb{F}_2$. Our implementation
uses Gray-code index order, as SciPy does. This reorders each initial
power-of-two block without changing its points. The Sobol' sequence is a
$(t,d)$-sequence in base $2$ \citep{Niederreiter1}, so an initial block of
$2^k$ points is a net when $k\ge t$. The quality parameter depends on the
dimension and the direction numbers. We use the direction numbers of
Joe and Kuo~\citep{JoeKuo2008}, chosen to improve two-dimensional projections,
as distributed with SciPy.

\subsection{Randomised quasi-Monte Carlo}

A randomisation of a low-discrepancy rule makes each point uniformly
distributed on the cube while keeping the structure that spreads the set
evenly. The estimator is then unbiased, and independent randomisations of the
same rule have a spread that estimates the error without the exact value.

Owen's scrambling \citep{Owen95,Owen97_1,Owen97_2} permutes each base-$b$ digit
of each coordinate by a random permutation that depends on all the preceding
digits. It maps a $(t,k,d)$-net to a $(t,k,d)$-net with probability one, and for
scrambled nets and sufficiently smooth integrands the RMSE is
$O(n^{-3/2+\epsilon})$ \citep{Owen97_2}. This theorem concerns the specified
scramble and integrand regularity, and is not a general rate for every
randomisation used below. For Sobol' points we use a cheaper random linear matrix
scramble followed by a random digital shift, the scheme of SciPy. The digits
beyond the thirtieth are drawn at random. The Faure sequence receives the same kind of
randomisation in base $b$, a random linear matrix scramble of its digits followed
by a random digital shift. The Halton sequence receives SciPy's
digit-permutation scramble, and the rank-1 lattice rules, built by the
component-by-component algorithm \citep{NuyensCools2006}, a uniform random shift
modulo one.

\section{Experimental design}
\label{sec:design}

\subsection{The basket grid}
\label{sec:basket-grid}

The basket family has $d=5$ independent assets, $\rho_{ij}=\delta_{ij}$, each
with $S_i(0)=100$, and the risk-free rate is $r=5\%$. The maturities,
volatilities and strikes are
\begin{align*}
T &\in \{0.15,\; 0.25,\; 0.5,\; 1,\; 2\},\\
\sigma &= 0.21 + 0.05\,(k-1), & k &= 1,\dots,10,\\
K_j &= 93 + j, & j &= 1,\dots,10,
\end{align*}
so the family holds $5\times 10\times 10=500$ contracts. All five assets of a
contract share the same volatility. Every contract is priced at every sample
size, and as in Algorithm~\ref{alg:protocol} the variance of a randomised rule,
or the squared error of a deterministic one, is averaged over the $500$
contracts before the square root is taken. The benchmark also includes three
deterministic rules, Sobol' points with the first $256$ points skipped and $n$ a
power of two, Faure points with the first $625$ points skipped and $n$ a power
of five, and the aligned Sobol' net that skips no points.

\subsection{The Asian contract and its path construction}

The Asian contracts have $S(0)=100$, $r=5\%$ and maturity $T=1$ year, with
$m=121$ monitoring dates $t_i=iT/m$, a spacing of about three days, so the
integral has dimension $121$. The discretisations $m=64$ and $m=256$ check the
results against the number of dates. The comparisons of path constructions in
Sections~\ref{sec:decoupling} and~\ref{sec:multiasset} fit each exponent to a
family of nine contracts, with volatilities of $15$, $20$ and $25$ per cent and
strikes $90$, $100$ and $110$, averaging the variance over the family as in
Algorithm~\ref{alg:protocol}. Their ANOVA diagnostics, the experiments of
Sections~\ref{sec:preint}, \ref{sec:multipreint} and~\ref{sec:windows}, and
Appendix~\ref{app:median} use the central contract, with $\sigma=0.2$ and
$K=100$. The errors of the Asian contracts are estimated from the spread of the
randomisations, except in Appendix~\ref{app:median}, which uses a reference
price.

The values of the Brownian motion at the $m$ dates can be generated from the $m$
input coordinates in several ways. The order matters because the quality of
multidimensional projections depends on which coordinates are used. The
incremental construction draws the increments in time order. The Brownian
bridge \citep{CaflischMorokoffOwen1997} draws the terminal value $W(t_m)$ from the
first coordinate. It then fills the date in the middle of the index range,
conditionally on the two endpoints, and continues breadth first, halving the
index ranges until every date is filled. Conditionally on $W(t_a)$ and $W(t_b)$
the value at a date $t_c$ between them is Gaussian,
\[
  W(t_c)\mid W(t_a),W(t_b)\sim\mathcal{N}\Bigl(\frac{(t_b-t_c)W(t_a)+(t_c-t_a)W(t_b)}{t_b-t_a},\;\frac{(t_c-t_a)(t_b-t_c)}{t_b-t_a}\Bigr).
\]
The principal-component construction writes the vector $(W(t_1),\dots,W(t_m))$
as $\sum_j\sqrt{\lambda_j}\,v_jz_j$, with the eigenvalues $\lambda_j$ of its
covariance in decreasing order, the corresponding eigenvectors $v_j$, and
independent standard normal variables $z_j$. The three constructions produce the
same joint Brownian distribution and differ in how path variation is assigned
to the input coordinates. The Brownian bridge and principal components place
large-scale path variation early.

\section{Protocol and diagnostics}
\label{sec:protocol}

The reference-price metric of Section~\ref{sec:figure-of-merit} requires known
contract values. Independent randomisations instead allow us to estimate the
RMSE and the uncertainty of its fitted slope. We draw $R$ independent
randomisations of the same rule, such as a scrambled net or a randomly shifted
lattice. Writing $\hat\alpha^{(r)}(n)$ for
the estimate from randomisation $r$ with $n$ points,
\begin{equation}
  \widehat{\mathrm{RMSE}}(n)
  = \Bigl[\tfrac{1}{R-1}\textstyle\sum_{r=1}^{R}
      \bigl(\hat\alpha^{(r)}(n)-\bar{\alpha}(n)\bigr)^2\Bigr]^{1/2},
  \qquad
  \bar\alpha(n)=\tfrac{1}{R}\textstyle\sum_{r=1}^{R}\hat\alpha^{(r)}(n).
\end{equation}
For a rule with uniform point marginals, each randomisation is unbiased for
the integral of the implemented integrand, so its root mean square error
equals its standard deviation. The spread estimates this error without the
exact value. Its square is unbiased for the variance, but the square root need
not be unbiased. This is the error of one randomisation of $n$ points. The
mean of $R$ independent randomisations has standard error
$\sigma(n)/\sqrt{R}$ and costs $Rn$ integrand evaluations. With fixed $R$,
both have the same population slope. Finite-precision sampling and numerical
payoff evaluation can introduce additional error. The spread cannot detect
a common bias or an incorrect integrand, which is why independent correctness
checks remain necessary.

The exponent $\beta$ in $\mathrm{RMSE}(n)\approx c\,n^{\beta}$ is estimated by
ordinary least squares of $\log_{10}\widehat{\mathrm{RMSE}}(n)$ on
$\log_{10}n$ over a declared set of sample sizes, which we call the fit window.
Unless stated otherwise the window is $n=2^{10},2^{11},\dots,2^{20}$. The
fitted slope $\hat\beta$ describes how the error decays over that window. It is not an estimate of an asymptotic rate, and it can change with the window.

For a nested rule all sample sizes of one randomisation are prefixes of a
single point set, so the values along one curve are dependent. The interval for
$\hat\beta$ therefore resamples whole randomisations to preserve this dependence.
For scrambled Sobol' points a prefix of length $2^k$ is itself a scrambled
$(t,k,d)$-net for $k\ge t$. In the dimensions of the basket benchmark,
the Halton and Faure constructions use bases other than two, so their
power-of-two prefixes do not have this base-two net guarantee. The
rank-1 lattice is the only rule
that is not nested, because the component-by-component algorithm
\citep{NuyensCools2006} builds its generating vector for a single $n$. It is
rebuilt at every sample size, randomisation $r$ uses the same random shift at
every sample size, and the lattice is reported only up to $n=2^{13}$, where the
construction is still affordable.

Four intervals are computed from the same randomisations. Let
$\hat\beta^{*}_1,\dots,\hat\beta^{*}_{N_B}$ be the slopes refitted after
drawing the $R$ randomisations with replacement, $q_a$ their $a$-quantile, and
$\hat\beta_{(1)},\dots,\hat\beta_{(R)}$ the slopes refitted with one
randomisation left out, with mean $\bar\beta_{(\cdot)}$.
\begin{itemize}
\item The percentile interval is $[q_{0.025},\,q_{0.975}]$.
\item The basic interval is $[2\hat\beta-q_{0.975},\,2\hat\beta-q_{0.025}]$.
\item The bias-corrected and accelerated (BCa) interval
  \citep{Efron1987better} adjusts the two quantile levels, with a bias
  correction from the share of the
  $\hat\beta^{*}_b$ below $\hat\beta$ and an acceleration from the skewness of
  the $\hat\beta_{(r)}$.
\item The jackknife-$t$ interval is
  $\hat\beta\pm t_{R-1,\,0.975}\,\widehat{\mathrm{se}}_{J}$, where
  $\widehat{\mathrm{se}}_{J}^{2}=\frac{R-1}{R}\sum_{r}
  \bigl(\hat\beta_{(r)}-\bar\beta_{(\cdot)}\bigr)^{2}$. It is the analogue, for
  a fitted slope, of the Student interval that L'Ecuyer et
  al.~\citep{LEcuyerNakayamaOwenTuffin2023} found more reliable than bootstrap
  intervals for the mean of randomised quasi-Monte Carlo estimates.
\end{itemize}
The Student quantile does not make the jackknife interval exact at finite $R$,
even when the individual replicate errors are Gaussian.

\begin{algorithm}[H]
\caption{Convergence exponent with a randomisation-level interval}
\label{alg:protocol}
\begin{algorithmic}[1]
\Require nested randomised rule $Q$ (scrambled net, randomised sequence or
  plain Monte Carlo); contracts $1,\dots,C$; sizes $n_1<\dots<n_L$;
  randomisations $R$; bootstrap draws $N_B$
\Ensure $\hat\beta$ and a nominal $95\%$ interval
\For{$r=1,\dots,R$}
  \State draw one randomisation of $Q$ with $n_L$ points
  \For{$\ell=1,\dots,L$}
    \State $\hat\alpha^{(r)}_{\ell,c}\gets$ mean over the first $n_\ell$ points,
      for each contract $c$
      \Comment{prefixes of one point set}
  \EndFor
\EndFor
\For{$\ell=1,\dots,L$}
  \State $M_\ell\gets\bigl[\tfrac1C\sum_c \operatorname{Var}_r
    \hat\alpha^{(r)}_{\ell,c}\bigr]^{1/2}$
    \Comment{sample variance over randomisations, $\chi$ never used}
\EndFor
\State $\hat\beta\gets$ slope of $\log_{10}M$ on $\log_{10}n$
\State $\hat\beta^{*}_1,\dots,\hat\beta^{*}_{N_B}\gets$ slopes after resampling
  whole randomisations, with replacement
\State $\hat\beta_{(1)},\dots,\hat\beta_{(R)}\gets$ slopes with randomisation $r$
  deleted
\State \Return $\hat\beta$ and the BCa interval built from
  $\hat\beta$, the $\hat\beta^{*}_b$ and the $\hat\beta_{(r)}$
\end{algorithmic}
\end{algorithm}

For a family of $C$ contracts, $M_\ell$ in Algorithm~\ref{alg:protocol}
averages the variance over the family before the square root is taken, so
that one exponent describes the family, and contracts with a larger variance
weigh more in it. When two rules are compared, the difference of their slopes has an interval of
its own, which resamples pairs of randomisations when the two rules share their
seeds.

\subsection{What the interval estimates}
\label{sec:target}

An interval from the protocol refers to a rule, an integrand and a fit window
$\mathcal{W}=\{n_1<\dots<n_L\}$ together. Let $e_\ell$ be the error
$\hat\alpha^{(r)}(n_\ell)-\alpha$ of one randomisation with $n_\ell$ points,
where $\alpha$ is the integral of the implemented integrand, and let
$\sigma_\ell^2=\mathbb{E}\,e_\ell^2$. The quantity estimated is the
least-squares slope of $\log\sigma_\ell$ on $\log n_\ell$ over the window,
\begin{equation}
  \beta_{\mathcal{W}}=\sum_{\ell=1}^{L}w_\ell\log\sigma_\ell,
  \qquad
  w_\ell=\frac{x_\ell-\bar x}{\sum_{k=1}^{L}(x_k-\bar x)^2},
  \qquad
  x_\ell=\log n_\ell,
\end{equation}
where $\bar x$ is the mean of $x_1,\dots,x_L$. Unqualified logarithms in the
formulas below are natural logarithms. Changing the base of both axes does
not change the slope. If $\sigma_\ell=c\,n_\ell^{\beta}$ for every $\ell$, then
$\beta_{\mathcal{W}}=\beta$. Otherwise $\beta_{\mathcal{W}}$ depends on the
window. The estimate $\hat\beta$ is the same
sum with $s_\ell=\widehat{\mathrm{RMSE}}(n_\ell)$ in place of $\sigma_\ell$.
The logarithms require positive sample variances. A zero-variance sample or
resample does not define a finite slope.
For the asymptotic statements below, extend the slope by any fixed finite
value on such samples. Their probability tends to zero under the stated
assumptions. The implementation instead reports an undefined fit explicitly.

\begin{proposition}[proof in Appendix~\ref{app:asymptotics}]
\label{prop:slope}
Fix a window with $L\ge2$ distinct sample sizes. Suppose that the $R$
error vectors are independent and identically distributed, that
$\mathbb{E}\,e_\ell=0$, and that $\sigma_\ell^2>0$ and $\mathbb{E}\,e_\ell^4<\infty$ for
$\ell=1,\dots,L$. Then, as $R\to\infty$,
$\sqrt{R}\,(\hat\beta-\beta_{\mathcal{W}})$ converges in distribution to
$\mathcal{N}(0,v_{\mathcal{W}})$, where
\[
  v_{\mathcal{W}}=\sum_{\ell=1}^{L}\sum_{k=1}^{L}
  \frac{w_\ell w_k\operatorname{Cov}\bigl(e_\ell^2,e_k^2\bigr)}
       {4\,\sigma_\ell^2\sigma_k^2}.
\]
Let two rules $A$ and $B$ satisfy the same conditions and share each
randomisation, so that the pairs of their error vectors are independent and
identically distributed over $r$. Then
\[
  \sqrt{R}\,\bigl(\hat\beta_B-\hat\beta_A
  -\beta_{\mathcal{W},B}+\beta_{\mathcal{W},A}\bigr)
\]
converges in distribution to $\mathcal{N}(0,v_A+v_B-2c_{AB})$, where
\[
  c_{AB}=\sum_{\ell=1}^{L}\sum_{k=1}^{L}
  \frac{w_\ell w_k\operatorname{Cov}\bigl(e_{A,\ell}^2,e_{B,k}^2\bigr)}
       {4\,\sigma_{A,\ell}^2\sigma_{B,k}^2}.
\]
If the two rules are randomised independently, $c_{AB}=0$.
\end{proposition}

For a fixed family of $C$
contracts sharing each randomisation, set
$q_\ell=C^{-1}\sum_c e_{\ell,c}^2$ and
$\sigma_\ell^2=\mathbb{E}q_\ell$. The same formulas hold with
$\operatorname{Cov}(q_\ell,q_k)$ in place of
$\operatorname{Cov}(e_\ell^2,e_k^2)$, provided every contract has zero-mean
errors and finite fourth moments. This corresponds to the pooled sample
variances in Algorithm~\ref{alg:protocol}, with positive pooled population
variance at every sample size. For a single contract a
diagonal term of $v_{\mathcal{W}}$ equals $w_\ell^2(\kappa_\ell-1)/4$, where
$\kappa_\ell=\mathbb{E}\,e_\ell^4/\sigma_\ell^4$ is the kurtosis of the error.
The other terms depend on how the squared errors at different sample sizes vary
together. They need not be small, because for a nested rule every sample size
is a prefix of the same point set, so the marginal kurtosis alone does not
determine the uncertainty of a slope. For the Gaussian family of
Section~\ref{sec:known-answer} every term is explicit,
\[
  \frac{\operatorname{Cov}\bigl(e_\ell^2,e_k^2\bigr)}{4\,\sigma_\ell^2\sigma_k^2}
  =\frac{\min(n_\ell,n_k)}{2\max(n_\ell,n_k)},
\]
because the estimate at $n$ is the mean of the first $n$ of a sequence of
independent Gaussian draws. Over the window $2^{10},\dots,2^{20}$ this gives an
asymptotic standard error of $0.137/\sqrt{R}$ for $\hat\beta$. The mean jackknife standard
error measured in the calibration is within four per cent of this value at
$R=32$ and within two per cent at $R=128$ and $R=512$.

Under the conditions of the proposition, and when $v_{\mathcal{W}}>0$, the
jackknife satisfies $R\widehat{\mathrm{se}}_J^2\to v_{\mathcal{W}}$ in
probability. The conditional bootstrap distribution of
$\sqrt{R}(\hat\beta^*-\hat\beta)$ consistently estimates the limiting
distribution of $\sqrt{R}(\hat\beta-\beta_{\mathcal{W}})$, because the slope is
a smooth function of empirical first and second moments
\citep{ShaoTu1995}. Thus the four intervals are first-order asymptotically
valid, with bootstrap quantiles understood as exact or with $N_B\to\infty$.
The fixed $N_B=2000$ adds Monte Carlo error to their endpoints.
For paired rules the corresponding conclusion requires
$v_A+v_B-2c_{AB}>0$ and resampling whole pairs. It also applies to independently
randomised rules treated as pairs. A zero contrast variance, for example for
identical rules, requires separate treatment.

The conditions fail in one of the calibration families. For the grid family of
Section~\ref{sec:known-answer} with $1/2<p<3/4$, the error has a finite variance
and an infinite fourth moment. The probability that its absolute value exceeds
$t$ decays like $t^{-1/(1-p)}$, so $e_\ell^2$ is in the domain of attraction of
a stable law of index $1/(2-2p)$, which lies between $1$ and $2$. The
fluctuations of $s_\ell^2$ are then of order $R^{1-2p}$ rather than $R^{-1/2}$
and are not asymptotically normal. The ordinary bootstrap can be inconsistent
for such means in the domain of attraction of a non-Gaussian stable law
\citep{Athreya1987}. The proposition therefore supplies no coverage guarantee
here. This does not by itself prove inconsistency for every slope or slope
difference, because dependence across sample sizes can produce cancellations.
When the fourth moment is finite but large, as at $p=1$, the proposition
applies without specifying how many randomisations the normal approximation
needs.

\subsection{Calibration on integrands with a known exponent}
\label{sec:known-answer}

We measure finite-sample coverage of the four intervals on integrands whose
exponent is known exactly, with replicate errors ranging from Gaussian to
heavy-tailed.

The first family is plain Monte Carlo on an integrand whose value at a uniform
point is a standard normal variable. With nested prefixes the estimates at all
$n$ are the means of the first $n$ of one sequence of independent $N(0,1)$
draws, which we simulate exactly through independent block sums, so
$\mathrm{RMSE}(n)=n^{-1/2}$ exactly, $\beta=-1/2$, and the replicates are
Gaussian at every $n$. This is a favourable case for a bootstrap interval.

The second family keeps the replicate errors non-Gaussian at every sample
size, as randomised quasi-Monte Carlo errors can be. For $p>1/2$ let
\begin{equation}
  f_p(x)=\sum_{h\neq 0}|h|^{-p}\,e^{2\pi\mathrm{i}hx},\qquad x\in[0,1),
\end{equation}
which defines an $L^2$ function. For $p\le1$ its value at the singular
endpoint can be assigned arbitrarily without changing the integral. Apply the
randomly shifted equispaced rule
$Q_n(U)=\frac1n\sum_{k=0}^{n-1}f_p(\{U+k/n\})$, with one uniform shift $U$ per
randomisation shared by all $n$. Averaging over the grid keeps only the
Fourier modes whose frequency is a multiple of $n$, and since the coefficients
are a power of $|h|$ the error rescales exactly for almost every shift $U$,
\begin{equation}
  Q_n(U)-\int_0^1 f_p(x)\,dx = n^{-p}\,f_p(\{nU\}).
\end{equation}
The RMSE is therefore $\sqrt{2\zeta(2p)}\,n^{-p}$ at every $n$ and
$\beta=-p$, with no pre-asymptotic regime. At every $n$ the error multiplied by $n^p$ has the
distribution of $f_p$ at a uniform point, so the replicates never become
Gaussian, and $p$ sets how heavy their tails are. At $p=2$ the function is a
bounded polynomial, $f_2(x)=2\pi^2(x^2-x+\tfrac16)$, and the kurtosis of the
error is $15/7$. At $p=1$ it has a logarithmic singularity,
$f_1(x)=-2\log(2\sin\pi x)$, all its moments are finite, and the kurtosis is
$57/5=11.4$. For $1/2<p<3/4$ its variance is finite and its fourth moment is
not, so at $p=0.6$ the kurtosis is infinite. For $p<1$ we evaluate $f_p$
through Hurwitz's formula for the zeta function.

Each family is run with $R=32$, $128$ and $512$ randomisations over the window
$n=2^{10},\dots,2^{20}$, with $N_B=2000$ bootstrap draws. Coverage is the share of $2000$ independent repetitions of the whole protocol
whose interval contains the true exponent (Table~\ref{tab:calibration}). With $2000$
repetitions the standard error of a coverage near $95\%$ is about half a
percentage point. The estimates for different $R$ reuse the same random stream, so
they are not independent of one another.

\begin{table}[H]
\centering
{\footnotesize
\setlength{\tabcolsep}{4pt}
\begin{tabular}{lcrccccc}
\toprule
Integrand & $\beta$ & $R$ & BCa & Percentile & Basic & Jackknife-$t$ & Bias of $\hat\beta$ \\
\midrule
Gaussian & $-0.5$ & 32 & 92.7$^{\ast}$ & 93.3$^{\ast}$ & 92.3$^{\ast}$ & 95.3 {\scriptsize($+10\%$)} & $0.0000$ \\
 & $-0.5$ & 128 & 94.5 & 94.5 & 94.3 & 95.3 {\scriptsize($+3\%$)} & $-0.0004$ \\
 & $-0.5$ & 512 & 95.3 & 95.4 & 95.2 & 95.5 {\scriptsize($+1\%$)} & $0.0000$ \\
\midrule
Grid, $p=2$ & $-2$ & 32 & 95.3 & 94.2 & 95.0 & 95.8 {\scriptsize($+5\%$)} & $+0.0007$ \\
 & $-2$ & 128 & 95.0 & 94.5 & 94.8 & 95.0 {\scriptsize($+2\%$)} & $+0.0005$ \\
 & $-2$ & 512 & 94.8 & 94.9 & 94.9 & 95.1 {\scriptsize($+1\%$)} & $+0.0001$ \\
\midrule
Grid, $p=1$ & $-1$ & 32 & 87.6$^{\ast}$ & 89.8$^{\ast}$ & 83.0$^{\ast}$ & 92.3$^{\ast}$ {\scriptsize($+24\%$)} & $-0.0002$ \\
 & $-1$ & 128 & 90.8$^{\ast}$ & 92.4$^{\ast}$ & 90.1$^{\ast}$ & 94.0$^{\ast}$ {\scriptsize($+10\%$)} & $-0.0002$ \\
 & $-1$ & 512 & 91.8$^{\ast}$ & 92.9$^{\ast}$ & 92.7$^{\ast}$ & 94.0$^{\ast}$ {\scriptsize($+4\%$)} & $-0.0001$ \\
\midrule
Grid, $p=0.6$ & $-0.6$ & 32 & 82.7$^{\ast}$ & 89.5$^{\ast}$ & 69.8$^{\ast}$ & 92.2$^{\ast}$ {\scriptsize($+46\%$)} & $+0.0026$ \\
 & $-0.6$ & 128 & 82.3$^{\ast}$ & 91.0$^{\ast}$ & 72.7$^{\ast}$ & 92.6$^{\ast}$ {\scriptsize($+37\%$)} & $+0.0027$ \\
 & $-0.6$ & 512 & 82.1$^{\ast}$ & 90.8$^{\ast}$ & 74.5$^{\ast}$ & 93.0$^{\ast}$ {\scriptsize($+33\%$)} & $+0.0036$ \\
\bottomrule
\end{tabular}}

\caption{Coverage in per cent of four nominal $95\%$ intervals for the fitted
exponent, each from $2000$ repetitions of the whole protocol on an integrand
whose exponent $\beta$ is known. A star marks a coverage whose Wilson interval
excludes $95\%$. The jackknife-$t$ column also gives its mean width relative
to BCa. The last column is the mean error of the fitted slope.}
\label{tab:calibration}
\end{table}

With Gaussian replicates the three bootstrap intervals undercover at $R=32$,
where BCa covers $92.7\%$. At $R=128$ and $512$ all four are within sampling error of
$95\%$. Jackknife-$t$ is compatible with nominal coverage already at $R=32$, where it is
$10\%$ wider than BCa. On the bounded integrand, $p=2$, every interval is
compatible with nominal coverage at each tested $R$.

The higher-kurtosis families behave differently. At $p=1$ no interval reaches
$95\%$ even with $R=512$, where the jackknife-$t$ interval covers $94.0\%$ and
BCa $91.8\%$. At $p=0.6$ more randomisations do not help. BCa covers $82.1\%$ at $R=512$,
no more than at $R=32$, and jackknife-$t$, the best of the four, covers at most
$93.0\%$ while being a third to nearly a half wider
(Table~\ref{tab:calibration}). In every family
the observed mean error of the fitted slope is at most
$0.0036$ in absolute value.

Coverage differs substantially across these families, which differ in both
their marginal distributions and their dependence across sample sizes. This
pattern is consistent with Proposition~\ref{prop:slope}. Its conditions hold at
$p=1$, where the coverage of BCa rises with $R$, and fail at $p=0.6$, where it
does not. Among
the four constructions, jackknife-$t$ gives the most reliable coverage in these
examples, partly because it is wider, and it still undercovers in the
heavy-tailed cases. The calibration integrands are one-dimensional, their
errors follow an exact power law, and each slope belongs to a single integrand,
whereas the error curves of the pricing contracts can bend and their slopes can
pool several contracts.

The sample kurtosis of the replicates places the contracts of
Sections~\ref{sec:preint} and~\ref{sec:multipreint} on the scale of the
calibration integrands (Table~\ref{tab:kurtosis}). For each contract and sample
size
we compute the sample kurtosis of the $R=512$ replicates, and report its median
and maximum over the eleven sample sizes and, for the single-asset Asian
contracts, over the three discretisations. For the calibration integrands the
median and maximum are taken over $400$ samples of $512$ draws. The
comparison has to be made at equal sample size, because with an infinite
fourth moment the sample kurtosis grows with the sample.

\begin{table}[H]
\centering
{\footnotesize
\begin{tabular}{lcc}
\toprule
Replicates & Median & Max \\
\midrule
\multicolumn{3}{l}{\emph{Calibration integrands}} \\
Gaussian & 3.0 & 4.6 \\
Grid, $p=2$ & 2.1 & 2.7 \\
Grid, $p=1$ & 9.8 & 57.6 \\
Grid, $p=0.6$ & 54.8 & 499.8 \\
\midrule
\multicolumn{3}{l}{\emph{Contracts, plain RQMC}} \\
Digital Asian, $m=64,121,256$ & 2.9 & 3.3 \\
Barrier Asian, $m=64,121,256$ & 3.0 & 3.3 \\
Basket, arithmetic, $\rho=0.0$ & 3.2 & 3.7 \\
Basket, digital, $\rho=0.0$ & 3.2 & 3.3 \\
Basket, arithmetic, $\rho=0.3$ & 3.5 & 7.1 \\
Basket, digital, $\rho=0.3$ & 2.9 & 3.3 \\
Basket, arithmetic, $\rho=0.9$ & 6.3 & 10.7 \\
Basket, digital, $\rho=0.9$ & 2.8 & 3.1 \\
\midrule
\multicolumn{3}{l}{\emph{Contracts, after preintegration}} \\
Digital Asian, $m=64,121,256$ & 4.7 & 7.0 \\
Barrier Asian, $m=64,121,256$ & 2.9 & 3.4 \\
Basket, arithmetic, $\rho=0.0$ & 3.8 & 5.3 \\
Basket, digital, $\rho=0.0$ & 3.2 & 3.7 \\
Basket, arithmetic, $\rho=0.3$ & 3.7 & 4.5 \\
Basket, digital, $\rho=0.3$ & 3.4 & 4.0 \\
Basket, arithmetic, $\rho=0.9$ & 4.6 & 6.0 \\
Basket, digital, $\rho=0.9$ & 4.2 & 4.6 \\
\bottomrule
\end{tabular}}

\caption{Sample kurtosis of the replicates, which is about $3$ for a Gaussian
sample. Calibration integrands: $512$ draws, median and maximum over $400$
samples. Contracts: $R=512$ randomisations at each sample size from $2^{10}$ to
$2^{20}$, median and maximum over the sample sizes and, for the single-asset
Asian contracts, over $m=64$, $121$ and $256$.}
\label{tab:kurtosis}
\end{table}

The median kurtosis of the replicates of these contracts lies between $2.8$ and
$4.7$, except for the plain arithmetic basket at $\rho=0.9$ with $6.3$, and every
median is below the $9.8$ of the grid with $p=1$. The maxima of the contracts and of the
calibration integrands are taken over different numbers of samples and are not
compared.

Sample kurtosis is a warning diagnostic. It can understate tail risk and does
not establish interval coverage. For the calibration integrands its
median is $9.8$ at $p=1$, where the kurtosis of the error is $11.4$, and $54.8$
at $p=0.6$, where it is infinite. A sample of $512$ can also miss rare large
errors altogether. Moreover, by Proposition~\ref{prop:slope} the uncertainty of
a slope depends on how the errors at different sample sizes vary together,
which the marginal kurtosis does not describe. The pricing experiments
use $R=512$ and BCa intervals unless marked otherwise.
At $R=512$ the calibration found BCa and jackknife-$t$ equally reliable when the
replicates were close to Gaussian. For
the contracts with the highest kurtosis, such as the arithmetic basket at
$\rho=0.9$, the BCa intervals may be too narrow.

\subsection{Calibration of intervals on a difference}
\label{sec:difference-calibration}

We measure the coverage of intervals on the difference of two slopes on pairs
of the integrands of Section~\ref{sec:known-answer} whose exponents differ by a
known amount $\Delta\beta$, the exponent of the second integrand minus that of
the first. The grids with $p=2$ and $p=1$ differ by $+1$, the grids with $p=1$
and $p=0.6$ by $+0.4$, and the Gaussian family and the grid with $p=1$ by
$-0.5$. Under the paired design the two integrands use the same uniform shift
in each randomisation, as two rules run on the same seeds share their
randomisation. Under the independent design each side draws its own. The
Gaussian family has no shift to share and appears only in the independent
design. Every repetition is analysed twice, once resampling pairs of
randomisations and once resampling the two sides apart, so that both analyses
see the same draws. The window, the numbers of randomisations, the $2000$
repetitions and the bootstrap draws are those of Section~\ref{sec:known-answer},
and the intervals are the percentile and jackknife-$t$ intervals of the
difference.

\begin{table}[H]
\centering
{\footnotesize
\setlength{\tabcolsep}{3.5pt}
\begin{tabular}{llcrccccc}
\toprule
 & & & & \multicolumn{2}{c}{Paired analysis} & \multicolumn{2}{c}{Unpaired analysis} & Width \\
\cmidrule(lr){5-6}\cmidrule(lr){7-8}
Integrands & Design & $\Delta\beta$ & $R$ & Percentile & Jackknife-$t$ & Percentile & Jackknife-$t$ & ratio \\
\midrule
$p=2$ vs $p=1$ & paired & $+1$ & 32 & 88.7$^{\ast}$ & 91.1$^{\ast}$ & 99.2$^{\ast}$ & 99.1$^{\ast}$ & $0.67$ \\
 &  & $+1$ & 128 & 92.2$^{\ast}$ & 93.8$^{\ast}$ & 98.8$^{\ast}$ & 98.7$^{\ast}$ & $0.74$ \\
 &  & $+1$ & 512 & 93.5$^{\ast}$ & 94.2 & 98.7$^{\ast}$ & 98.7$^{\ast}$ & $0.77$ \\
\midrule
$p=2$ vs $p=1$ & independent & $+1$ & 32 & 89.0$^{\ast}$ & 91.8$^{\ast}$ & 89.5$^{\ast}$ & 92.2$^{\ast}$ & $1.01$ \\
 &  & $+1$ & 128 & 93.2$^{\ast}$ & 94.3 & 93.7$^{\ast}$ & 94.4 & $1.00$ \\
 &  & $+1$ & 512 & 93.5$^{\ast}$ & 94.0$^{\ast}$ & 93.5$^{\ast}$ & 94.2 & $1.00$ \\
\midrule
$p=1$ vs $p=0.6$ & paired & $+0.4$ & 32 & 88.6$^{\ast}$ & 91.5$^{\ast}$ & 100.0$^{\ast}$ & 100.0$^{\ast}$ & $0.44$ \\
 &  & $+0.4$ & 128 & 90.0$^{\ast}$ & 92.5$^{\ast}$ & 99.9$^{\ast}$ & 99.8$^{\ast}$ & $0.62$ \\
 &  & $+0.4$ & 512 & 90.5$^{\ast}$ & 92.7$^{\ast}$ & 99.4$^{\ast}$ & 99.0$^{\ast}$ & $0.76$ \\
\midrule
$p=1$ vs $p=0.6$ & independent & $+0.4$ & 32 & 89.4$^{\ast}$ & 93.0$^{\ast}$ & 89.6$^{\ast}$ & 92.8$^{\ast}$ & $1.01$ \\
 &  & $+0.4$ & 128 & 92.7$^{\ast}$ & 94.5 & 92.3$^{\ast}$ & 94.5 & $1.00$ \\
 &  & $+0.4$ & 512 & 91.5$^{\ast}$ & 93.5$^{\ast}$ & 91.6$^{\ast}$ & 93.7$^{\ast}$ & $1.00$ \\
\midrule
Gaussian vs $p=1$ & independent & $-0.5$ & 32 & 91.0$^{\ast}$ & 94.1 & 91.3$^{\ast}$ & 93.7$^{\ast}$ & $1.01$ \\
 &  & $-0.5$ & 128 & 93.4$^{\ast}$ & 94.3 & 93.3$^{\ast}$ & 94.2 & $1.00$ \\
 &  & $-0.5$ & 512 & 94.4 & 95.0 & 94.4 & 94.9 & $1.00$ \\
\bottomrule
\end{tabular}}

\caption{Coverage in per cent of nominal $95\%$ intervals for a difference of
two exponents, each from $2000$ repetitions on integrands whose difference
$\Delta\beta$ is known. Each repetition is analysed by resampling pairs of
randomisations and by resampling the two sides apart. A star marks a coverage
whose Wilson interval excludes $95\%$. The last column is the mean width of the
paired jackknife-$t$ interval over that of the unpaired one.}
\label{tab:difference}
\end{table}

When the two integrands share their randomisation, resampling pairs gives
intervals $0.44$ to $0.77$ times as wide as resampling the sides apart, which
overcovers because it ignores the positive dependence that the shared shift
creates between the two slopes (Table~\ref{tab:difference}). By
Proposition~\ref{prop:slope}, under its finite-moment and nondegeneracy
conditions, ignoring a negative covariance would instead underestimate the
asymptotic variance. When the two sides are drawn independently, the two
analyses agree in coverage and width, so treating independent draws as pairs
costs nothing in these examples.

Resampling pairs preserves the dependence under both designs, but it does not
remedy undercoverage from non-Gaussian errors. When neither side is the grid
with $p=0.6$, the paired jackknife-$t$ interval covers between $93.8\%$ and
$95.0\%$ at $R=128$ and $512$. When one side is that grid, it covers about as
much as that grid alone. The percentile interval covers less than jackknife-$t$
in every paired row, and the fitted differences have small mean errors. These cases have large
differences between integrands of different smoothness.

Six further cases use the same protocol (Table~\ref{tab:small-difference}).
For a nontrivial null comparison, both sides use $p=1$, with either independent
shifts or $U_B=\{U_A+1/3\}$. The latter gives distinct curves with the same
population exponent. For small effects, $p=1$ is compared with $p=0.98$ or
$p=0.95$, giving $\Delta\beta=+0.02$ or $+0.05$, with shared or independent
shifts. Identical rules on identical shifts would instead give an identically
zero contrast and would not test interval calibration.

\begin{table}[H]
\centering
{\footnotesize
\setlength{\tabcolsep}{3.5pt}
\begin{tabular}{llcrccccc}
\toprule
 & & & & \multicolumn{2}{c}{Paired jackknife-$t$} & \multicolumn{2}{c}{Unpaired jackknife-$t$} & Width \\
\cmidrule(lr){5-6}\cmidrule(lr){7-8}
Integrands & Design & $\Delta\beta$ & $R$ & Coverage & Excludes $0$ & Coverage & Excludes $0$ & ratio \\
\midrule
$p=1$ vs $p=1$ & shift $+1/3$ & $+0.00$ & 32 & 94.9 & 5.1 & 93.5$^{\ast}$ & 6.5 & $1.06$ \\
 &  & $+0.00$ & 128 & 94.3 & 5.7 & 93.5$^{\ast}$ & 6.6 & $1.04$ \\
 &  & $+0.00$ & 512 & 95.7 & 4.3 & 94.6 & 5.4 & $1.03$ \\
\midrule
$p=1$ vs $p=1$ & independent & $+0.00$ & 32 & 93.7$^{\ast}$ & 6.3 & 93.5$^{\ast}$ & 6.5 & $1.01$ \\
 &  & $+0.00$ & 128 & 94.0 & 5.9 & 94.2 & 5.8 & $1.00$ \\
 &  & $+0.00$ & 512 & 94.8 & 5.2 & 94.7 & 5.3 & $1.00$ \\
\midrule
$p=1$ vs $p=0.98$ & paired & $+0.02$ & 32 & 91.0$^{\ast}$ & 99.5 & 100.0$^{\ast}$ & 0.0 & $0.02$ \\
 &  & $+0.02$ & 128 & 91.4$^{\ast}$ & 99.9 & 100.0$^{\ast}$ & 0.0 & $0.03$ \\
 &  & $+0.02$ & 512 & 93.4$^{\ast}$ & 100.0 & 100.0$^{\ast}$ & 0.0 & $0.04$ \\
\midrule
$p=1$ vs $p=0.98$ & independent & $+0.02$ & 32 & 94.7 & 6.0 & 94.3 & 6.1 & $1.01$ \\
 &  & $+0.02$ & 128 & 94.3 & 8.1 & 94.4 & 8.4 & $1.00$ \\
 &  & $+0.02$ & 512 & 95.3 & 15.8 & 95.5 & 15.7 & $1.00$ \\
\midrule
$p=1$ vs $p=0.95$ & paired & $+0.05$ & 32 & 91.0$^{\ast}$ & 98.9 & 100.0$^{\ast}$ & 0.1 & $0.06$ \\
 &  & $+0.05$ & 128 & 92.0$^{\ast}$ & 99.8 & 100.0$^{\ast}$ & 0.4 & $0.09$ \\
 &  & $+0.05$ & 512 & 93.7$^{\ast}$ & 100.0 & 100.0$^{\ast}$ & 79.1 & $0.11$ \\
\midrule
$p=1$ vs $p=0.95$ & independent & $+0.05$ & 32 & 94.1 & 10.7 & 94.4 & 10.1 & $1.01$ \\
 &  & $+0.05$ & 128 & 94.8 & 21.1 & 94.7 & 20.9 & $1.00$ \\
 &  & $+0.05$ & 512 & 95.5 & 60.8 & 95.4 & 61.3 & $1.00$ \\
\bottomrule
\end{tabular}}

\caption{Null and small differences, each from $2000$ repetitions. Coverage
and the frequency of excluding zero are percentages for nominal $95\%$
jackknife-$t$ intervals. Excluding zero is a false positive under the null and
a detection under an alternative. A star marks a coverage whose Wilson interval
excludes $95\%$. The width ratio is paired over unpaired. The offset design
uses $U_B=\{U_A+1/3\}$.}
\label{tab:small-difference}
\end{table}

At $R=512$ the paired analysis is close to nominal under both null designs, and
so is coverage for the small alternatives with independent shifts, whose
intervals however exclude zero in only $15.8\%$ and $60.8\%$ of repetitions
(Table~\ref{tab:small-difference}). Sharing the shift between these similar
integrands produces much stronger cancellation. The paired intervals shrink to
$0.04$ and $0.11$ of the unpaired width and exclude zero in every repetition,
yet cover the true differences in only $93.4\%$ and $93.7\%$. High detection
frequency can thus coexist with undercoverage. These differences are of the size of the barrier comparisons,
but their deliberately strong coupling does not reproduce the
dependence of the pricing estimators.

\subsection{The ANOVA decomposition}
\label{sec:anova}

Any $f\in L^2([0,1]^d)$ can be written uniquely, up to equality almost everywhere,
as a sum of $2^d$ mutually orthogonal terms
\begin{equation}
  f(\mathbf{x})=\sum_{u\subseteq\{1,\dots,d\}} f_u(\mathbf{x}_u),
  \qquad
  \sigma^2(f)=\sum_{u\neq\varnothing}\sigma_u^2,
\end{equation}
where $f_\varnothing=\mathbb{E}f$, and each nonconstant $f_u$ depends only on
the coordinates in $u$ and integrates to zero in each of them. Assume
$\sigma^2(f)>0$. The \emph{closed} Sobol' index of a set $u$ is the share of
variance explained by those coordinates jointly,
\begin{equation}
  S^{\mathrm{cl}}_u=\frac{1}{\sigma^2(f)}\sum_{\varnothing\ne v\subseteq u}\sigma_v^2
  =\frac{\operatorname{Var}\bigl(\mathbb{E}[f\mid \mathbf{x}_u]\bigr)}{\sigma^2(f)},
\end{equation}
and the \emph{truncation dimension} at level $\gamma$ is the length of the
shortest leading block of coordinates that reaches it,
$d_T(\gamma)=\min\{k: S^{\mathrm{cl}}_{\{1,\dots,k\}}\ge \gamma\}$
\citep{CaflischMorokoffOwen1997,WangFang2003}. We use $\gamma=0.99$. We write
$S_j=S^{\mathrm{cl}}_{\{j\}}$ for the first-order indices. Their sum is one
exactly when $f$ is additive, so $\sum_j S_j$ is the share of variance that
involves no interactions.

Sobol' indices and effective dimensions have been computed for option-pricing
integrands before
\citep{Bianchetti2015pricing,ScoleriBianchettiKucherenko2021,Kucherenko2011effective}.

The indices are estimated with the pick-freeze estimator of Janon et
al.~\citep{JanonEtAl2014} on scrambled Sobol' points. The truncation dimension
is found by bisection on $k$, which assumes that the estimated closed index
increases with $k$. The search is repeated eight times with $2^{18}$ points
each, and the reported value is the median of the eight values, rounded up. In most
cases the eight values agree to within one coordinate. Where the truncation
dimension is above two hundred they can differ by up to eighteen, and those
entries have appreciable numerical variability. These observed ranges are
not confidence bounds on the true truncation dimension. The sums of first-order
indices use five repetitions with $2^{20}$ points each. Their reported
uncertainty is the estimated standard error across those repetitions.

The two samples in the pick-freeze identity must have independent uniform
components within each matched row. We obtain them from the two halves of one
scrambled $2d$-dimensional net, which also gives joint low-discrepancy structure.
Separately and independently scrambled $d$-dimensional nets satisfy the
probabilistic identity too, though their joint quadrature properties can differ.
Reusing a scramble can violate the required independence. Additive test
functions, whose first-order indices sum to one, provide a useful check.

\section{The protocol applied to the basket benchmark}
\label{sec:corrections}

\begin{figure}[H]
\centering
\includegraphics[width=0.86\textwidth]{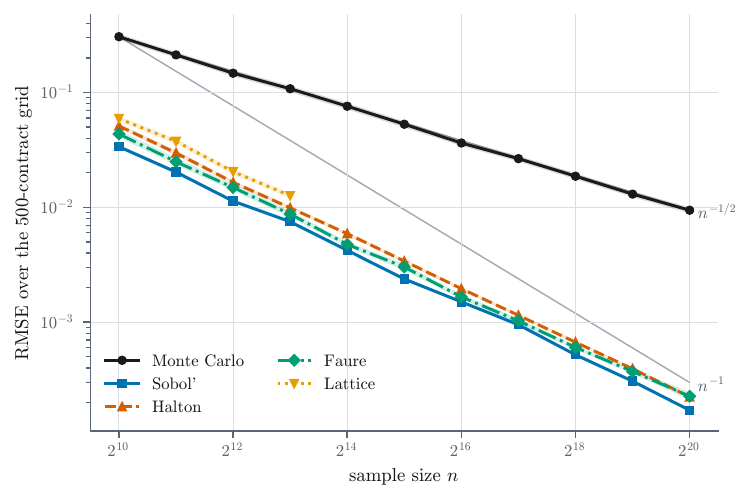}
\caption{Root mean square error against the exact prices for the randomised
rules of Table~\ref{tab:basket}, with nominal $95\%$ pointwise percentile
bootstrap bands. The two grey lines have
slopes $-1/2$ and $-1$ and start from the Monte Carlo error at the first sample
size.}
\label{fig:convergence}
\Description{Log-log error curves for Monte Carlo and four randomised
quasi-Monte Carlo rules, with pointwise uncertainty bands. Sobol' has the
smallest observed error, although Halton has the steepest fitted slope.}
\end{figure}

The protocol is applied to the family of $500$ geometric basket contracts of
Section~\ref{sec:basket-grid}, at $R=512$ and with $n$ up to $2^{20}$
(Table~\ref{tab:basket} and Figure~\ref{fig:convergence}).

\begin{table}[H]
\centering
\begin{tabular}{lccc}
\toprule
Rule & Exponent {\scriptsize [95\%]} & Error at largest $n$ & $n$ \\
\midrule
Monte Carlo & $-0.502$ {\scriptsize $[-0.513,\,-0.491]$} & $9.44\times 10^{-3}$ & $2^{20}$ \\
\midrule
Sobol' (scrambled) & $-0.755$ {\scriptsize $[-0.762,\,-0.748]$} & $1.71\times 10^{-4}$ & $2^{20}$ \\
Halton (scrambled) & $-0.780$ {\scriptsize $[-0.787,\,-0.774]$} & $2.24\times 10^{-4}$ & $2^{20}$ \\
Faure (scrambled) & $-0.761$ {\scriptsize $[-0.779,\,-0.749]$} & $2.27\times 10^{-4}$ & $2^{20}$ \\
Rank-1 lattice, CBC & $-0.756$ {\scriptsize $[-0.791,\,-0.722]$} & $1.26\times 10^{-2}$ & $2^{13}$ \\
Sobol', first 256 skipped$^{\dagger}$ & $-0.778$ & $1.27\times 10^{-4}$ & $2^{20}$ \\
Sobol', aligned net$^{\dagger}$ & $-0.870$ & $1.64\times 10^{-4}$ & $2^{20}$ \\
Faure, first 625 skipped$^{\dagger}$ & $-0.979$ & $2.56\times 10^{-3}$ & 78125 \\
\bottomrule
\end{tabular}

\caption{Convergence exponents on the geometric basket family. The Monte Carlo
row, whose exponent is known to be $-1/2$, checks the protocol. The three rows marked $\dagger$ are
deterministic rules, which give one error per sample size and no randomisation,
so their slopes are descriptive and carry no interval. All intervals are BCa
over $R=512$ randomisations. The error column is the root mean square error
at the largest sample size, from the spread of the randomisations or, for the
deterministic rules, from the exact values.}
\label{tab:basket}
\end{table}

The Monte Carlo estimate is $-0.502$, and its interval contains $-1/2$.

The Halton points have the steepest fitted slope and the Sobol' points the
smallest error, and the marginal intervals of the two are disjoint. The Faure
interval overlaps both. The lattice interval is wider. Its
shorter window, with only four sample sizes up to $2^{13}$, contributes to its
uncertainty and prevents a comparison over the same range. It contains the
point estimates of the other three rules. A faster rate and a smaller
error at a given budget are different claims, and here they rank the rules
differently. Among the randomised rules, Sobol' has the smallest error at every
shared sample size tested.

For the deterministic Sobol' rule, skipping the first $256$ points breaks the
alignment that makes a block of $2^k$
consecutive points a net \citep{Owen2020FirstPoint}, and over this window the
aligned rule has the steeper fitted slope. The skipped rule nevertheless has
the smaller error at every sample size. Skipping also removes the first point,
the origin, which the inverse normal transform maps far into the lower tail in
every coordinate. The share of the difference due to each
effect cannot be separated with these two rules. A scrambled net keeps the alignment and moves the origin with
probability one.

\section{The test contract as a ridge function}
\label{sec:ridge}

In the lognormal form of Section~\ref{sec:n_simu}, the only random part of the
geometric basket's exponent is $\tfrac{1}{d}\sum_{i=1}^d\sigma_i W_i(T)$, a
single linear functional of the $d$ normal variables. The payoff is therefore
a ridge function with at most one effective direction, for any volatilities,
correlation matrix and number of assets. Adding assets does not increase its
intrinsic dimension.

The closed index $S^{\mathrm{cl}}_{1:k}$ of the first $k$ coordinates and the
truncation dimension $d_T(\gamma)$ of Section~\ref{sec:anova} measure this under
the identity ordering of the coordinates and under a single orthogonal rotation
that aligns the first coordinate with the ridge direction
(Table~\ref{tab:effdim}). The rotation leaves the price
unchanged and reduces the truncation dimension from $5$ to $1$. Under the
identity ordering the first-order indices sum to $0.770$, so the coordinates
interact, and after the rotation they sum to $1.000$.

\begin{table}[H]
\centering
\begin{tabular}{lccccccc}
\toprule
Coordinate ordering & \multicolumn{5}{c}{$S^{\mathrm{cl}}_{1:k}$} & $d_T(0.99)$ & $\sum_j S_j$ \\
\cmidrule(lr){2-6}
 & $k{=}1$ & $2$ & $3$ & $4$ & $5$ & & \\
\midrule
Identity ordering & 0.154 & 0.329 & 0.525 & 0.747 & 1.000 & 5 & 0.770 \\
Rotated onto the ridge & 1.000 & 1.000 & 1.000 & 1.000 & 1.000 & 1 & 1.000 \\
\bottomrule
\end{tabular}

\caption{Effective dimension of one geometric basket contract with $d=5$,
$T=1$, volatility $31\%$ and $K=100$. The closed indices are single estimates
from $2^{18}$ points. Under the identity ordering the closed index increases with $k$. After one
rotation the first coordinate carries all the variance.}
\label{tab:effdim}
\end{table}

A dimension sweep on this contract cannot isolate the effect of increasing
intrinsic dimension. The payoff remains a ridge, although its distribution
and coordinate effective dimension can change. Hoyt and
Owen~\citep{HoytOwen2019} study the mean dimension of ridge
functions in general.

\section{Effective dimension and smoothness}
\label{sec:decoupling}

\subsection{The contract family}
\label{sec:contracts}

We use three payoffs on the discretely monitored arithmetic average
$\Bar{S}_{\mathrm{ari}}$ of Section~\ref{sec:asian-contract}, all discounted at
$e^{-rT}$ and monitored at the same $m$ dates. $K$ is the strike and $B$ the
barrier level, set at $130$ against $S(0)=100$. The barrier is an upper
knock-out on the asset price, not on its average.
\begin{align}
  \text{arithmetic} &: && (\Bar{S}_{\mathrm{ari}}-K)^+, \\
  \text{digital}    &: && \mathds{1}\{\Bar{S}_{\mathrm{ari}}>K\}, \\
  \text{barrier}    &: && (\Bar{S}_{\mathrm{ari}}-K)^+\,
                          \mathds{1}\Bigl\{\max_{1\le i\le m} S(t_i)<B\Bigr\}.
\end{align}
The arithmetic payoff is continuous, with a kink at the strike. The digital
payoff jumps across a single level set of the average. The barrier payoff has
the kink of the call and, in addition, a jump that depends on the running
maximum of the path, a functional of every monitoring date. The exponents are
fitted over the family of nine contracts of Section~\ref{sec:design} as in
Algorithm~\ref{alg:protocol}, and the truncation dimensions and first-order
indices are computed for its central contract (Table~\ref{tab:asian}).

Low effective dimension alone need not give a fast rate, particularly for
discontinuous payoffs \citep{Owen2003dimension,WangTan2013,HeWang2015discontinuous}.
The three Asian payoffs agree with that account. Their central-contract
diagnostics and pooled slopes describe them without isolating the effects of
dimension and smoothness.

\begin{table}[H]
\centering
\footnotesize
\begin{tabular}{llcccc}
\toprule
Payoff & Construction & $m$ & Exponent {\scriptsize [95\% BCa]} & $d_T(0.99)$ & $\sum_j S_j$ \\
\midrule
Arithmetic & Incremental & 121 & $-0.557$ {\scriptsize $[-0.565,\,-0.550]$} & 99 & $0.751 \pm 0.027$ \\
Arithmetic & Brownian bridge & 121 & $-0.708$ {\scriptsize $[-0.714,\,-0.703]$} & 7 & $0.898 \pm 0.003$ \\
Arithmetic & PCA & 121 & $-0.939$ {\scriptsize $[-0.956,\,-0.925]$} & 2 & $0.994 \pm 0.000$ \\
Digital & Incremental & 121 & $-0.509$ {\scriptsize $[-0.515,\,-0.503]$} & 116 & $0.590 \pm 0.055$ \\
Digital & Brownian bridge & 121 & $-0.556$ {\scriptsize $[-0.560,\,-0.551]$} & 52 & $0.670 \pm 0.012$ \\
Digital & PCA & 121 & $-0.573$ {\scriptsize $[-0.576,\,-0.569]$} & 6 & $0.886 \pm 0.004$ \\
Barrier & Incremental & 121 & $-0.507$ {\scriptsize $[-0.515,\,-0.499]$} & 121 & $0.159 \pm 0.034$ \\
Barrier & Brownian bridge & 121 & $-0.541$ {\scriptsize $[-0.548,\,-0.535]$} & 118 & $0.412 \pm 0.037$ \\
Barrier & PCA & 121 & $-0.534$ {\scriptsize $[-0.541,\,-0.527]$} & 119 & $0.603 \pm 0.072$ \\
\midrule
Arithmetic & Incremental & 256 & $-0.559$ {\scriptsize $[-0.567,\,-0.552]$} & 208 & $0.780 \pm 0.045$ \\
Arithmetic & Brownian bridge & 256 & $-0.707$ {\scriptsize $[-0.713,\,-0.702]$} & 7 & $0.892 \pm 0.008$ \\
Arithmetic & PCA & 256 & $-0.932$ {\scriptsize $[-0.954,\,-0.918]$} & 2 & $0.994 \pm 0.001$ \\
Digital & Incremental & 256 & $-0.501$ {\scriptsize $[-0.507,\,-0.495]$} & 245 & $0.598 \pm 0.061$ \\
Digital & Brownian bridge & 256 & $-0.558$ {\scriptsize $[-0.563,\,-0.554]$} & 53 & $0.681 \pm 0.037$ \\
Digital & PCA & 256 & $-0.579$ {\scriptsize $[-0.583,\,-0.575]$} & 6 & $0.887 \pm 0.021$ \\
Barrier & Incremental & 256 & $-0.496$ {\scriptsize $[-0.503,\,-0.489]$} & 255 & $0.283 \pm 0.076$ \\
Barrier & Brownian bridge & 256 & $-0.546$ {\scriptsize $[-0.552,\,-0.539]$} & 242 & $0.530 \pm 0.076$ \\
Barrier & PCA & 256 & $-0.537$ {\scriptsize $[-0.544,\,-0.530]$} & 249 & $0.600 \pm 0.028$ \\
\bottomrule
\end{tabular}

\caption{Asian contracts under three path constructions. Exponents are fitted
over the family of nine contracts, with BCa intervals at $R=512$. $d_T(0.99)$ is
the truncation dimension and $\sum_j S_j$ the sum of the first-order indices of
the central contract, and a sum near one indicates a nearly additive
integrand. The $\pm$ values are estimated standard errors of the first-order
sum across five independent numerical repetitions.}
\label{tab:asian}
\end{table}

For the continuous arithmetic payoff, lower measured dimension accompanies
a steeper fitted slope. At $m=121$, going from the
incremental construction to principal components lowers the truncation
dimension from $99$ to $2$ and steepens the exponent from $-0.557$, near the
Monte Carlo rate, to $-0.939$, near $n^{-1}$, with the Brownian bridge in
between.

The digital payoff has a low truncation dimension under principal components
but a much slower fitted rate. Its truncation
dimension is $6$ and its exponent $-0.573$, while the arithmetic payoff under the
Brownian bridge has truncation dimension $7$ and exponent $-0.708$. The two
integrands have almost the same effective dimension, and their exponents differ
by more than a tenth with intervals far apart, at $m=256$ as at $m=121$. Reducing the
digital payoff's truncation dimension from $116$ to $6$ improves its exponent
by less than a tenth, which is consistent with the fact that a rotation of the
coordinates does not remove a jump.

For the barrier payoff, none of the three constructions concentrates the
variance in a few coordinates. Its truncation dimension stays close to the nominal dimension, its exponent
within five hundredths of the Monte Carlo rate, and the first-order indices sum
to at most $0.603$, so a large share of the variance lies in interactions.

\section{Smoothing by preintegration}
\label{sec:preint}

Preintegration integrates out one variable analytically, which under suitable
regularity conditions leaves a smoother
function of the remaining $m-1$ variables for quasi-Monte Carlo
\citep{GriewankKuoLeoveySloan2018}. Under the
principal-component construction the leading factor has strictly positive
loadings, so the running average is increasing in the leading Gaussian
coordinate $z_1=\Phi^{-1}(x_1)$. The indicator
$\mathds{1}\{\Bar{S}_{\mathrm{ari}}>K\}$ therefore switches exactly once, at the
threshold $z_1^{*}$ where $\Bar{S}_{\mathrm{ari}}=K$. We find $z_1^{*}$ by a
Newton iteration safeguarded by bisection, and
$\mathbb{E}[\mathds{1}\{\Bar{S}_{\mathrm{ari}}>K\}\mid x_2,\dots,x_m]=1-\Phi(z_1^{*})$.
The implementation uses a finite bracket and iteration budget. Its root and
Gaussian-tail evaluations are numerical approximations to this formula.

The monotonicity is a hypothesis of the method, and the implementation checks
that every leading loading is positive. Gilbert, Kuo and
Sloan~\citep{GilbertKuoSloan2021} show that without it the preintegrated
function can fail to be smooth. Monotonicity alone does not guarantee a smooth
boundary for the barrier payoff. Under principal components every
path value has the form $S(t_i)=b_i\,e^{c_i z_1}$, where $c_i>0$ is a constant
and $b_i>0$ depends only on the remaining coordinates. The running maximum is therefore
increasing in $z_1$, and the knock-out level is crossed exactly once, at a
threshold $z_B$. The payoff is supported on the interval $(z_1^{*},z_B)$, and
its conditional expectation is again a Gaussian integral in closed form. The
monotonicity condition holds, but the running maximum is not a globally smooth
boundary function. The smoothness theorem therefore does not apply merely
because the crossing is unique. The threshold $z_B$ is a minimum over the
monitoring dates and can have kinks where the binding date changes. These
can persist in the conditional expectation, which is zero wherever
$z_B\le z_1^{*}$.

Both payoffs are conditioned on the leading factor, but they differ in more
than the regularity of the resulting integrands, so their rate changes do not
isolate the effect of smoothness.

\begin{table}[H]
\centering
\begin{tabular}{lccc}
\toprule
Payoff & $m$ & Plain RQMC & After preintegration \\
\midrule
Digital & 64 & $-0.580$ {\scriptsize $[-0.591,\,-0.569]$} & $-1.056$ {\scriptsize $[-1.074,\,-1.040]$} \\
Barrier & 64 & $-0.536$ {\scriptsize $[-0.548,\,-0.524]$} & $-0.556$ {\scriptsize $[-0.566,\,-0.544]$} \\
Digital & 121 & $-0.574$ {\scriptsize $[-0.585,\,-0.563]$} & $-1.058$ {\scriptsize $[-1.074,\,-1.042]$} \\
Barrier & 121 & $-0.533$ {\scriptsize $[-0.546,\,-0.520]$} & $-0.558$ {\scriptsize $[-0.571,\,-0.546]$} \\
Digital & 256 & $-0.574$ {\scriptsize $[-0.585,\,-0.563]$} & $-1.071$ {\scriptsize $[-1.086,\,-1.056]$} \\
Barrier & 256 & $-0.540$ {\scriptsize $[-0.552,\,-0.528]$} & $-0.553$ {\scriptsize $[-0.565,\,-0.542]$} \\
\bottomrule
\end{tabular}

\caption{Convergence exponent of the central Asian contract before and after
integrating out the leading principal component. Intervals are BCa at
$R=512$.}
\label{tab:preint}
\end{table}

The digital exponent moves past $-1$ at $m=64$, $121$ and $256$, an improvement
of about $0.5$ (Table~\ref{tab:preint}). This agrees with the
$O(n^{-1+\epsilon})$ rate that He~\citep{He2019rate} proves for conditional
randomised quasi-Monte Carlo under smoothness conditions. The barrier's exponent
changes by at most three hundredths. Conditioning removed the jump in $z_1$ and
left the kinks in the other coordinates.
The barrier's error nevertheless falls, through the constant rather than the
rate (Section~\ref{sec:windows}).

Barrier-specific schemes \citep{AchtsisCoolsNuyens2012} and other conditioning
directions \citep{Liu2022} may behave differently from conditioning on the
leading principal component. The choice of direction depends on the regularity
of the resulting conditional expectation as well as on the variance that the
direction carries.

\section{Effective dimension across assets and dates}
\label{sec:multiasset}

The contract is an Asian option on a basket of $d$ assets, each monitored at
the same $m$ dates, so the integral has dimension $d\,m$. With
$A_i$ the time average of asset $i$, we use three payoffs, the call
$(\tfrac1d\sum_iA_i-K)^+$ on the average of the $A_i$, its digital
$\mathds{1}\{\tfrac1d\sum_iA_i>K\}$, and the worst-of call
$(\min_iA_i-K)^+$. The assets in Table~\ref{tab:multiasset} are independent.

A construction now has to decide where each input coordinate goes. The
incremental construction is asset-major, so the first $m$ coordinates describe
the first asset completely before the second asset appears. The Brownian
bridge is level-major, so the first $d$ coordinates carry the $d$ terminal
values, the next $d$ the $d$ midpoints, and so on. An asset-major bridge would
put the second asset's terminal value at coordinate $m+1$, changing which
Sobol' projections carry the important path factors. The principal components
are those of the full $d\,m$ covariance, ordered by eigenvalue. This ordering
does not determine the construction completely when eigenvalues coincide
(Section~\ref{sec:correlation}). At $d=1$ the three constructions reduce to
those of Section~\ref{sec:decoupling}.

\begin{table}[H]
\centering
\footnotesize
\setlength{\tabcolsep}{4pt}
\begin{tabular}{llccccc}
\toprule
Construction & Payoff & \multicolumn{5}{c}{assets $d$ / dates $m$} \\
\cmidrule(lr){3-7}
 & & $1/64$ & $2/64$ & $3/64$ & $5/64$ & $5/121$ \\
\midrule
Incremental & Arithmetic & 53 / $-0.58$ & 113 / $-0.55$ & 175 / $-0.56$ & 299 / $-0.57$ & 565 / $-0.61$ \\
 & Digital & 62 / $-0.52$ & 125 / $-0.51$ & 188 / $-0.51$ & 316 / $-0.50$ & 596 / $-0.52$ \\
 & Worst-of & 53 / $-0.58$ & 117 / $-0.52$ & 181 / $-0.53$ & 310 / $-0.52$ & 585 / $-0.52$ \\
\midrule
Brownian bridge & Arithmetic & 7 / $-0.71$ & 14 / $-0.67$ & 20 / $-0.64$ & 33 / $-0.62$ & 34 / $-0.61$ \\
 & Digital & 43 / $-0.56$ & 87 / $-0.56$ & 130 / $-0.54$ & 218 / $-0.53$ & 266 / $-0.53$ \\
 & Worst-of & 7 / $-0.71$ & 16 / $-0.66$ & 37 / $-0.65$ & 78 / $-0.61$ & 79 / $-0.61$ \\
\midrule
PCA & Arithmetic & 2 / $-0.94$ & 3 / $-0.87$ & 4 / $-0.81$ & 6 / $-0.76$ & 6 / $-0.77$ \\
 & Digital & 6 / $-0.57$ & 11 / $-0.58$ & 16 / $-0.57$ & 26 / $-0.56$ & 26 / $-0.56$ \\
 & Worst-of & 2 / $-0.94$ & 4 / $-0.76$ & 6 / $-0.74$ & 14 / $-0.69$ & 14 / $-0.69$ \\
\bottomrule
\end{tabular}

\caption{The Asian option on a basket of $d$ independent assets, of dimension
$d\,m$. Each cell gives $d_T(0.99)$ for the central contract and the exponent
fitted over the family of nine contracts, at $R=512$ and $n$ up to $2^{20}$.
Principal components use the common eigenbasis of
Section~\ref{sec:correlation}.}
\label{tab:multiasset}
\end{table}

Under the incremental construction the truncation dimension stays close to the
nominal dimension, $299$ of $320$ for the arithmetic payoff at $d=5$ and $m=64$
(Table~\ref{tab:multiasset}).

Under the Brownian bridge and principal components the truncation dimension
grows with the number of assets and hardly changes with the number of dates.
Under principal components it rises from $2$ to $6$ for the arithmetic payoff
and from $6$ to $26$ for the digital between one and five assets, and at $d=5$
it is the same at $m=64$ and $m=121$ for all three payoffs. This observed growth with $d$ is specific to the tested contracts and
constructions, and gives no lower bound over all coordinate transformations.

The increase in truncation dimension accompanies a weaker exponent. Under
principal components the arithmetic exponent weakens from $-0.94$ with one asset
to $-0.76$ with five, and the worst-of to $-0.69$. At
$d=5$ the worst-of needs $14$ coordinates where the arithmetic payoff needs $6$.
A plausible reason is that its payoff depends on each asset separately, through
the minimum, while the leading factors favour the common direction.

Adding assets also changes the integrand, so dimension is not isolated as the
cause of the weaker exponent.

\subsection{PCA eigenbasis and asset correlation}
\label{sec:correlation}

With an equicorrelation $\rho$ between the assets the covariance becomes $C_\rho\otimes\Sigma$, with
$C_\rho=(1-\rho)I_d+\rho\mathbf{1}\mathbf{1}^{\top}$, instead of
$I_d\otimes\Sigma$. For $d>1$, this is a valid correlation matrix when
$-1/(d-1)\le\rho\le1$. The experiments use $0\le\rho<1$.
Its eigenvalues are products of the eigenvalues of the two
factors. $C_\rho$ has the eigenvalue $1+(d-1)\rho$ once, with the common
direction $\mathbf{1}/\sqrt{d}$ as eigenvector, and the eigenvalue $1-\rho$
with multiplicity $d-1$, on the contrasts between assets.

At $\rho=0$ every eigenvalue of $C_\rho$ equals one, so each eigenvalue
$\lambda_j$ of $\Sigma$ gives a $d$-dimensional eigenspace of the product, and
principal components alone do not fix a basis inside it. Two choices are
natural. The per-asset basis uses the vectors $e_i\otimes v_j$, one asset at a
time, where $v_j$ is the $j$-th eigenvector of $\Sigma$. The common basis uses
$\mathbf{1}/\sqrt{d}\otimes v_j$ first, followed by orthonormal Helmert
contrasts between the assets. In both, the eigenspaces are ordered by
eigenvalue. For $\rho>0$, each common-direction product
$(1+(d-1)\rho)\lambda_j$ contributes one eigenvector, and each contrast product
$(1-\rho)\lambda_j$ contributes $d-1$. Products from different temporal modes
can coincide, adding degeneracy. We use the same Helmert contrasts and a fixed
ordering of the product modes. Apart from the per-asset
rows at $\rho=0$ in Table~\ref{tab:correlation}, all principal-component results
use the common basis.

\begin{table}[H]
\centering
\setlength{\tabcolsep}{4pt}
\begin{tabular}{lllcc}
\toprule
$\rho$ & Basis & Payoff & $d_T(0.99)$ & Exponent {\scriptsize [95\% BCa]} \\
\midrule
$0.0$ & per-asset & Arithmetic & 8 & $-0.708$ {\scriptsize $[-0.714,\,-0.702]$} \\
$0.0$ & per-asset & Digital & 29 & $-0.571$ {\scriptsize $[-0.576,\,-0.565]$} \\
$0.0$ & per-asset & Worst-of & 14 & $-0.707$ {\scriptsize $[-0.715,\,-0.699]$} \\
\midrule
$0.0$ & common & Arithmetic & 6 & $-0.757$ {\scriptsize $[-0.764,\,-0.750]$} \\
$0.0$ & common & Digital & 26 & $-0.561$ {\scriptsize $[-0.566,\,-0.557]$} \\
$0.0$ & common & Worst-of & 14 & $-0.688$ {\scriptsize $[-0.694,\,-0.681]$} \\
\midrule
$0.3$ & common & Arithmetic & 6 & $-0.790$ {\scriptsize $[-0.799,\,-0.782]$} \\
$0.3$ & common & Digital & 18 & $-0.540$ {\scriptsize $[-0.545,\,-0.536]$} \\
$0.3$ & common & Worst-of & 11 & $-0.697$ {\scriptsize $[-0.703,\,-0.692]$} \\
\midrule
$0.6$ & common & Arithmetic & 6 & $-0.835$ {\scriptsize $[-0.845,\,-0.824]$} \\
$0.6$ & common & Digital & 14 & $-0.552$ {\scriptsize $[-0.556,\,-0.548]$} \\
$0.6$ & common & Worst-of & 7 & $-0.706$ {\scriptsize $[-0.712,\,-0.700]$} \\
\midrule
$0.9$ & common & Arithmetic & 2 & $-0.922$ {\scriptsize $[-0.937,\,-0.909]$} \\
$0.9$ & common & Digital & 10 & $-0.574$ {\scriptsize $[-0.578,\,-0.570]$} \\
$0.9$ & common & Worst-of & 7 & $-0.717$ {\scriptsize $[-0.723,\,-0.711]$} \\
\bottomrule
\end{tabular}

\caption{The same contracts at $d=5$ and $m=64$ under principal components,
against the correlation between the assets. At $\rho=0$ both eigenbases are
shown. The columns give the truncation dimension of the central contract and the
exponent fitted over the family, with its BCa interval.}
\label{tab:correlation}
\end{table}

The two bases span the same subspace in every block of $d$ leading
coordinates. Conditioning the payoff, viewed as a function of Gaussian inputs,
on its first $kd$ coordinates is
conditioning on its projection onto that subspace, so the closed index
$S^{\mathrm{cl}}_{1:kd}$ is the same in both bases for every $k$. The basis can
therefore move the truncation dimension only inside a block of $d$ coordinates.
At $\rho=0$ the two bases give truncation dimensions that differ by at most
three coordinates for every payoff, each pair within one block of five
(Table~\ref{tab:correlation}).

Inside a block the choice matters, and its direction depends on the payoff. At
$\rho=0$ the common basis steepens the arithmetic exponent from $-0.708$ to
$-0.757$, while the per-asset basis steepens the worst-of from $-0.688$ to
$-0.707$. Both pairs of marginal intervals are disjoint, and the two digital
intervals just overlap. A choice that principal components leave open therefore
changes the measured rate as well as the truncation dimension. These
are marginal intervals and not an interval on the difference. An exponent
reported for principal components on such a contract is incomplete without the
basis.

Correlation lowers the truncation dimension of every payoff, but not in
proportion to $\rho$. For the arithmetic payoff it stays at $6$ up to $\rho=0.6$
and drops to the single-asset value of $2$ only at $\rho=0.9$.

Correlation concentrates the variance without reducing the number of factors
that carry it. At $d=5$ and $\rho=0.9$ the eigenvalues of $C_\rho$ are $4.6$
and $0.1$, so there are still five independent Brownian factors. The share of
the variance carried by the first of them is what changes, and the truncation
dimension responds to that share.

The arithmetic exponent improves steadily with $\rho$, from $-0.757$ to
$-0.922$, and no two of the four intervals overlap. The digital exponent hardly
moves, although its truncation dimension falls from $26$ to $10$. As with the path rotations in
Section~\ref{sec:decoupling}, the reduction in dimension brings little
improvement in the digital payoff's rate.

\subsection{Smoothing the multi-asset contract}
\label{sec:multipreint}

Preintegration needs a coordinate in which the basket average is monotone. On this
contract the leading factor of the $d\,m$ covariance under the common basis is
such a coordinate. Its loadings are positive on every asset and every date, so
each asset's time average and the basket are increasing in it. The threshold is
unique, and the conditional expectation is a Gaussian integral of $d\,m$
exponential terms.

At $\rho=0$ the leading eigenspace has dimension $d$, and a general eigensolver
applied to the block-diagonal covariance $I_d\otimes\Sigma$ may return a leading
factor carried by a single asset. The basket is still non-decreasing in such a
factor, but the factor leaves the other $d-1$ assets unchanged, and our
implementation requires loadings that are positive on every asset and every
date. We therefore integrate out the common direction
$\mathbf{1}/\sqrt{d}\otimes v_1$, which lies in the same eigenspace, reproduces
the covariance equally well and loads equally on every asset. We did not compare
the two choices. At $\rho>0$ the leading eigenvalue $(1+(d-1)\rho)\lambda_1$ is simple,
and principal components select the common direction without a choice.

\begin{table}[H]
\centering
\setlength{\tabcolsep}{4pt}
\begin{tabular}{llcc}
\toprule
$\rho$ & Payoff & Plain RQMC & After preintegration \\
\midrule
$0.0$ & Arith & $-0.751$ {\scriptsize $[-0.763,\,-0.740]$} & $-0.981$ {\scriptsize $[-0.995,\,-0.967]$} \\
$0.0$ & Digital & $-0.560$ {\scriptsize $[-0.572,\,-0.548]$} & $-1.004$ {\scriptsize $[-1.015,\,-0.994]$} \\
$0.3$ & Arith & $-0.783$ {\scriptsize $[-0.799,\,-0.768]$} & $-0.989$ {\scriptsize $[-1.003,\,-0.975]$} \\
$0.3$ & Digital & $-0.528$ {\scriptsize $[-0.539,\,-0.518]$} & $-1.004$ {\scriptsize $[-1.015,\,-0.992]$} \\
$0.9$ & Arith & $-0.939$ {\scriptsize $[-0.964,\,-0.921]$} & $-1.022$ {\scriptsize $[-1.039,\,-1.005]$} \\
$0.9$ & Digital & $-0.567$ {\scriptsize $[-0.578,\,-0.557]$} & $-1.059$ {\scriptsize $[-1.073,\,-1.044]$} \\
\bottomrule
\end{tabular}

\caption{Convergence exponent of the central contract before and after
integrating out the leading factor under the common basis, at $d=5$ and
$m=64$. Intervals are BCa at
$R=512$.}
\label{tab:multipreint}
\end{table}

The digital exponent improves by about $0.5$ at every correlation and reaches
$-1$ (Table~\ref{tab:multipreint}). This is close to the single-asset gain
of Section~\ref{sec:preint}, so in these cases neither
the additional assets nor the correlation reduce what the smoothing achieves.

The arithmetic payoff gains less, particularly at higher correlation. Its exponent
improves by about a quarter at $\rho=0$, a fifth at $\rho=0.3$ and less than a
tenth at $\rho=0.9$, where the plain exponent is already $-0.939$. Its payoff
has a kink rather than a jump, so there was less to remove. Together with the
digital rows, this is consistent with a gain that is largest when the removed
coordinate carries a discontinuity. These
full-range figures average a gain that grows with $n$
(Section~\ref{sec:windows}).

The worst-of is not preintegrated, although its level is monotone in the
leading factor as well, as a minimum of increasing functions, so the threshold
exists. The asset
that attains the minimum changes along the coordinate, however, and the
conditional expectation becomes a sum over intervals whose endpoints are
crossings between sums of exponentials.

\section{Fit-window sensitivity and computational cost}
\label{sec:windows}

For the comparisons of Sections~\ref{sec:preint} and~\ref{sec:multipreint}, the
difference of exponents has an interval over five windows, the full range, the
range without its smallest sizes, the large-budget range, and the early and
late halves (Table~\ref{tab:windows}). The same five windows are
used for every contract. Each interval concerns one window and one contract,
with no simultaneous coverage guarantee across the collection.

Replicate $r$ of the plain and of the preintegrated rule uses the same seed. The
two Sobol' point sets then share the
scrambling matrices of their common coordinates but not their digital shifts,
and those coordinates feed different factors of the two constructions. The
estimates of the two rules are nearly uncorrelated. Their correlation lies
within two standard errors of zero at $126$ of the $132$ combinations of
contract and sample size. We resample the replicates in pairs, which requires
that the pairs are independent and identically distributed. This also permits
independence within a pair. Resampling the two rules separately changes each standard
error by less than five per cent and moves no interval across zero.

\begin{table}[H]
\centering
{\scriptsize
\setlength{\tabcolsep}{2.5pt}
\begin{tabular}{lrrrrr}
\toprule
 & \multicolumn{5}{c}{Difference in exponent over the window $\log_2 n$} \\
\cmidrule(lr){2-6}
Contract & $10$--$20$ & $12$--$20$ & $14$--$20$ & $10$--$15$ & $15$--$20$ \\
\midrule
Digital, $m{=}64$ & $-0.476$\,{\scriptsize $\pm 0.020$}\phantom{$^{\circ}$} & $-0.480$\,{\scriptsize $\pm 0.025$}\phantom{$^{\circ}$} & $-0.470$\,{\scriptsize $\pm 0.033$}\phantom{$^{\circ}$} & $-0.467$\,{\scriptsize $\pm 0.041$}\phantom{$^{\circ}$} & $-0.473$\,{\scriptsize $\pm 0.044$}\phantom{$^{\circ}$} \\
Digital, $m{=}121$ & $-0.484$\,{\scriptsize $\pm 0.020$}\phantom{$^{\circ}$} & $-0.484$\,{\scriptsize $\pm 0.026$}\phantom{$^{\circ}$} & $-0.477$\,{\scriptsize $\pm 0.035$}\phantom{$^{\circ}$} & $-0.484$\,{\scriptsize $\pm 0.042$}\phantom{$^{\circ}$} & $-0.491$\,{\scriptsize $\pm 0.041$}\phantom{$^{\circ}$} \\
Digital, $m{=}256$ & $-0.497$\,{\scriptsize $\pm 0.019$}\phantom{$^{\circ}$} & $-0.498$\,{\scriptsize $\pm 0.026$}\phantom{$^{\circ}$} & $-0.514$\,{\scriptsize $\pm 0.033$}\phantom{$^{\circ}$} & $-0.483$\,{\scriptsize $\pm 0.042$}\phantom{$^{\circ}$} & $-0.506$\,{\scriptsize $\pm 0.042$}\phantom{$^{\circ}$} \\
\midrule
Barrier, $m{=}64$ & $-0.020$\,{\scriptsize $\pm 0.016$}\phantom{$^{\circ}$} & $-0.021$\,{\scriptsize $\pm 0.020$}\phantom{$^{\circ}$} & $-0.011$\,{\scriptsize $\pm 0.030$}$^{\circ}$ & $-0.006$\,{\scriptsize $\pm 0.036$}$^{\circ}$ & $+0.000$\,{\scriptsize $\pm 0.036$}$^{\circ}$ \\
Barrier, $m{=}121$ & $-0.025$\,{\scriptsize $\pm 0.018$}\phantom{$^{\circ}$} & $-0.020$\,{\scriptsize $\pm 0.022$}$^{\circ}$ & $-0.012$\,{\scriptsize $\pm 0.030$}$^{\circ}$ & $-0.038$\,{\scriptsize $\pm 0.035$}\phantom{$^{\circ}$} & $-0.022$\,{\scriptsize $\pm 0.037$}$^{\circ}$ \\
Barrier, $m{=}256$ & $-0.014$\,{\scriptsize $\pm 0.017$}$^{\circ}$ & $-0.023$\,{\scriptsize $\pm 0.022$}\phantom{$^{\circ}$} & $-0.027$\,{\scriptsize $\pm 0.029$}$^{\circ}$ & $+0.016$\,{\scriptsize $\pm 0.032$}$^{\circ}$ & $-0.015$\,{\scriptsize $\pm 0.035$}$^{\circ}$ \\
\midrule
Digital, $\rho{=}0$ & $-0.444$\,{\scriptsize $\pm 0.016$}\phantom{$^{\circ}$} & $-0.454$\,{\scriptsize $\pm 0.021$}\phantom{$^{\circ}$} & $-0.458$\,{\scriptsize $\pm 0.031$}\phantom{$^{\circ}$} & $-0.421$\,{\scriptsize $\pm 0.039$}\phantom{$^{\circ}$} & $-0.455$\,{\scriptsize $\pm 0.039$}\phantom{$^{\circ}$} \\
Digital, $\rho{=}0.3$ & $-0.476$\,{\scriptsize $\pm 0.016$}\phantom{$^{\circ}$} & $-0.494$\,{\scriptsize $\pm 0.022$}\phantom{$^{\circ}$} & $-0.499$\,{\scriptsize $\pm 0.030$}\phantom{$^{\circ}$} & $-0.438$\,{\scriptsize $\pm 0.034$}\phantom{$^{\circ}$} & $-0.503$\,{\scriptsize $\pm 0.037$}\phantom{$^{\circ}$} \\
Digital, $\rho{=}0.9$ & $-0.492$\,{\scriptsize $\pm 0.019$}\phantom{$^{\circ}$} & $-0.492$\,{\scriptsize $\pm 0.025$}\phantom{$^{\circ}$} & $-0.480$\,{\scriptsize $\pm 0.033$}\phantom{$^{\circ}$} & $-0.490$\,{\scriptsize $\pm 0.038$}\phantom{$^{\circ}$} & $-0.474$\,{\scriptsize $\pm 0.041$}\phantom{$^{\circ}$} \\
\midrule
Arithmetic, $\rho{=}0$ & $-0.230$\,{\scriptsize $\pm 0.019$}\phantom{$^{\circ}$} & $-0.286$\,{\scriptsize $\pm 0.026$}\phantom{$^{\circ}$} & $-0.317$\,{\scriptsize $\pm 0.033$}\phantom{$^{\circ}$} & $-0.088$\,{\scriptsize $\pm 0.039$}\phantom{$^{\circ}$} & $-0.333$\,{\scriptsize $\pm 0.038$}\phantom{$^{\circ}$} \\
Arithmetic, $\rho{=}0.3$ & $-0.207$\,{\scriptsize $\pm 0.022$}\phantom{$^{\circ}$} & $-0.254$\,{\scriptsize $\pm 0.028$}\phantom{$^{\circ}$} & $-0.293$\,{\scriptsize $\pm 0.035$}\phantom{$^{\circ}$} & $-0.079$\,{\scriptsize $\pm 0.047$}\phantom{$^{\circ}$} & $-0.308$\,{\scriptsize $\pm 0.037$}\phantom{$^{\circ}$} \\
Arithmetic, $\rho{=}0.9$ & $-0.082$\,{\scriptsize $\pm 0.028$}\phantom{$^{\circ}$} & $-0.099$\,{\scriptsize $\pm 0.035$}\phantom{$^{\circ}$} & $-0.135$\,{\scriptsize $\pm 0.042$}\phantom{$^{\circ}$} & $-0.028$\,{\scriptsize $\pm 0.055$}$^{\circ}$ & $-0.143$\,{\scriptsize $\pm 0.051$}\phantom{$^{\circ}$} \\
\bottomrule
\end{tabular}

\medskip
\begin{tabular}{lrrrrrrrr}
\toprule
 & \multicolumn{2}{c}{Plain} & \multicolumn{2}{c}{Preintegrated} & \multicolumn{4}{c}{RMSE ratio, preintegrated / plain} \\
\cmidrule(lr){2-3}\cmidrule(lr){4-5}\cmidrule(lr){6-9}
Contract & $10$--$15$ & $15$--$20$ & $10$--$15$ & $15$--$20$ & $n{=}2^{10}$ & $n{=}2^{20}$ & min & max \\
\midrule
Digital, $m{=}64$ & $-0.590$ & $-0.588$ & $-1.057$ & $-1.062$ & $0.00647$ & $0.000240$ & $0.000240$ & $0.00647$ \\
Digital, $m{=}121$ & $-0.574$ & $-0.564$ & $-1.057$ & $-1.055$ & $0.00696$ & $0.000248$ & $0.000248$ & $0.00696$ \\
Digital, $m{=}256$ & $-0.590$ & $-0.568$ & $-1.073$ & $-1.074$ & $0.00761$ & $0.000233$ & $0.000233$ & $0.00761$ \\
\midrule
Barrier, $m{=}64$ & $-0.540$ & $-0.535$ & $-0.547$ & $-0.534$ & $0.138$ & $0.136$ & $0.115$ & $0.141$ \\
Barrier, $m{=}121$ & $-0.514$ & $-0.526$ & $-0.552$ & $-0.548$ & $0.168$ & $0.138$ & $0.138$ & $0.168$ \\
Barrier, $m{=}256$ & $-0.562$ & $-0.512$ & $-0.546$ & $-0.527$ & $0.140$ & $0.142$ & $0.130$ & $0.157$ \\
\midrule
Digital, $\rho{=}0$ & $-0.580$ & $-0.550$ & $-1.001$ & $-1.005$ & $0.0148$ & $0.000827$ & $0.000827$ & $0.0148$ \\
Digital, $\rho{=}0.3$ & $-0.555$ & $-0.496$ & $-0.993$ & $-1.000$ & $0.0116$ & $0.000425$ & $0.000425$ & $0.0116$ \\
Digital, $\rho{=}0.9$ & $-0.582$ & $-0.569$ & $-1.072$ & $-1.043$ & $0.00787$ & $0.000276$ & $0.000276$ & $0.00787$ \\
\midrule
Arithmetic, $\rho{=}0$ & $-0.872$ & $-0.662$ & $-0.959$ & $-0.994$ & $0.244$ & $0.0658$ & $0.0658$ & $0.273$ \\
Arithmetic, $\rho{=}0.3$ & $-0.888$ & $-0.688$ & $-0.967$ & $-0.997$ & $0.156$ & $0.0355$ & $0.0355$ & $0.156$ \\
Arithmetic, $\rho{=}0.9$ & $-0.998$ & $-0.867$ & $-1.026$ & $-1.010$ & $0.0781$ & $0.0422$ & $0.0422$ & $0.0781$ \\
\bottomrule
\end{tabular}

\medskip
\begin{tabular}{lccc}
\toprule
 & \multicolumn{2}{c}{Cost ratio, preintegrated / plain} & RMSE ratio \\
\cmidrule(lr){2-3}
Contract & same $n$ & $2^{19}$ against $2^{20}$ & $2^{19}$ against $2^{20}$ \\
\midrule
Digital, $m{=}64$ & $2.11$ & $1.05$ & $0.000531$ {\scriptsize $[0.000487,\,0.000579]$} \\
Digital, $m{=}121$ & $2.06$ & $1.03$ & $0.000505$ {\scriptsize $[0.000452,\,0.000566]$} \\
Digital, $m{=}256$ & $2.03$ & $1.01$ & $0.000467$ {\scriptsize $[0.000418,\,0.000518]$} \\
\midrule
Barrier, $m{=}64$ & $2.38$ & $1.19$ & $0.185$ {\scriptsize $[0.170,\,0.201]$} \\
Barrier, $m{=}121$ & $2.35$ & $1.17$ & $0.196$ {\scriptsize $[0.180,\,0.215]$} \\
Barrier, $m{=}256$ & $2.32$ & $1.16$ & $0.195$ {\scriptsize $[0.178,\,0.213]$} \\
\midrule
Digital, $\rho{=}0$ & $1.96$ & $0.98$ & $0.00137$ {\scriptsize $[0.00126,\,0.00149]$} \\
Digital, $\rho{=}0.3$ & $1.97$ & $0.99$ & $0.000814$ {\scriptsize $[0.000742,\,0.000891]$} \\
Digital, $\rho{=}0.9$ & $1.97$ & $0.98$ & $0.000553$ {\scriptsize $[0.000497,\,0.000610]$} \\
\midrule
Arithmetic, $\rho{=}0$ & $2.11$ & $1.06$ & $0.130$ {\scriptsize $[0.118,\,0.142]$} \\
Arithmetic, $\rho{=}0.3$ & $2.12$ & $1.06$ & $0.0721$ {\scriptsize $[0.0659,\,0.0792]$} \\
Arithmetic, $\rho{=}0.9$ & $2.11$ & $1.06$ & $0.0799$ {\scriptsize $[0.0711,\,0.0891]$} \\
\bottomrule
\end{tabular}
}

\caption{Preintegrated against plain rules over five fit windows, with $R=512$
randomisations paired by seed. Top: difference in exponent, preintegrated minus
plain, with the half-width of its jackknife-$t$ interval, marked $^{\circ}$ when
the interval contains zero. Middle: exponents over the early and late halves,
and the RMSE ratio, preintegrated over plain, at $n=2^{10}$ and $n=2^{20}$
together with its minimum and maximum over $n$. Bottom: cost ratios, the second estimated by assuming that cost is
proportional to $n$, and the RMSE of the preintegrated rule at $n=2^{19}$ over
that of the plain rule at $n=2^{20}$ with a percentile interval.
Rows labelled by $m$ are single-asset Asian options, rows labelled by $\rho$ the
basket with $d=5$ and $m=64$.}
\label{tab:windows}
\end{table}

The digital gain changes little with the window. It lies between $-0.514$ and
$-0.421$ in every window, at every discretisation and every correlation, and no
interval comes near zero. For this payoff preintegration improves the rate over
the whole observed range.

The barrier gain does depend on the window. Over the full range the difference
lies between $-0.025$ and $-0.014$, each discretisation has exactly one shorter
window whose interval excludes zero, and the sign of the difference depends on
the window. The RMSE ratio, preintegrated over plain, lies between $0.115$ and
$0.168$ at every sample size and discretisation. Conditioning therefore divides
the barrier's error by six to nine at every sample size, while its exponent
changes by less than four hundredths in every window.
Conditioning cannot increase the variance of a plain Monte Carlo estimate.
Under randomised quasi-Monte Carlo there is no such guarantee, and here the
error fell at every sample size although the conditioned function retains
kinks.

\begin{figure}[H]
\centering
\includegraphics[width=0.92\textwidth]{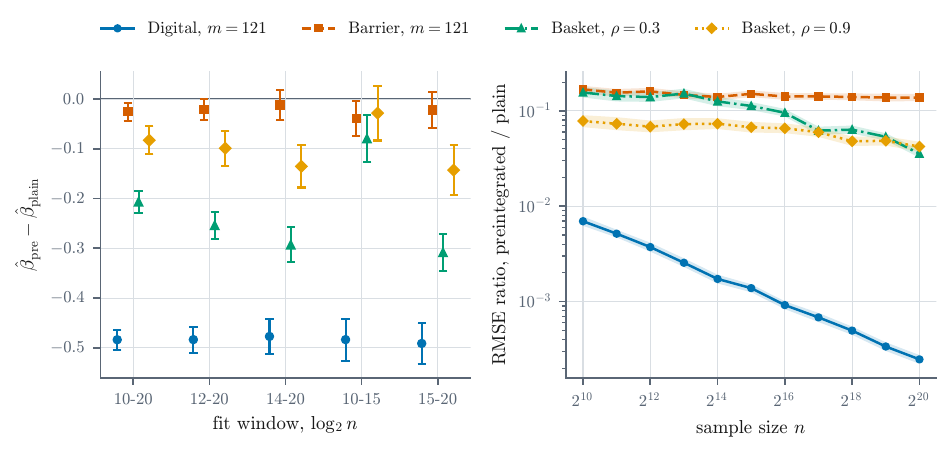}
\caption{Left: the difference in exponent over each fit window, with its
jackknife-$t$ interval. Right: the RMSE ratio at every $n$, with its percentile
pointwise band. The digital payoff gains in rate, the barrier mainly in error
magnitude, and the arithmetic basket's gain grows at larger sample sizes.}
\label{fig:windows}
\Description{Left panels show slope differences and jackknife intervals over
five fit windows. Right panels show preintegrated-to-plain error ratios with
pointwise bands. Digital payoffs improve in slope, barriers mainly in error
magnitude, and arithmetic baskets show a larger gain at larger sample sizes.}
\end{figure}

For the arithmetic basket the window matters most. At every
correlation its gain is roughly four to five times larger over the late half of
the window than over the early half, from $-0.088$ to $-0.333$ at $\rho=0$. The
preintegrated exponent changes little between the halves, while the plain
exponent slows, from $-0.872$ to $-0.662$ at $\rho=0$. The full-range exponents of Section~\ref{sec:multipreint}
average this change. The RMSE ratio at $\rho=0$ falls from $0.244$ at
$n=2^{10}$ to $0.0658$ at $n=2^{20}$.
In every window the gain is largest at $\rho=0$ and smallest at $\rho=0.9$.

These ratios compare equal numbers of points, and preintegration adds a root
solve for each point. One replicate of
the preintegrated rule, a nested curve up to $n=2^{20}$, costs between $1.96$
and $2.38$ times as much as one of the plain rule. The timings include point
generation, the payoff and the root solves, but not the one-off construction of
the preintegrator, and come from a single timed pass without warm-up. For the
single-asset contracts the timed plain rule also recomputes its principal
components in every replicate, so the cost ratios there are somewhat
understated. If the cost of a curve is proportional to its length, the
preintegrated rule at $n=2^{19}$ costs about as much as the plain rule at
$n=2^{20}$. At this budget preintegration still divides the barrier's error by
about five instead of seven, the arithmetic basket's by eight to fourteen and
the digital payoffs' by more than seven hundred (Table~\ref{tab:windows}). Timings depend on the hardware and on the
implementation, so these factors are specific to this setting.

\section{Conclusions}
\label{sec:conclusions-v2}

The protocol estimates a finite-window slope from independent randomisations
and resamples whole curves to construct its interval. Under the stated moment
and nondegeneracy conditions, these intervals are asymptotically valid as $R$
grows. Their finite-sample coverage depends on the error distribution. All four
constructions attain coverage compatible with $95\%$ for Gaussian errors at
$R=128$ and $512$, and jackknife-$t$ already at $R=32$. Jackknife-$t$ is the
most reliable of the four in these experiments, but all four undercover for the
two higher-kurtosis families even at $R=512$. In comparisons between rules,
resampling pairs preserves dependence and can substantially reduce interval
width. Strong coupling and high detection frequency do not ensure nominal coverage.
These calibrations use one-dimensional integrands with exact power-law errors,
so they do not establish coverage for the pricing estimators, for RMSE ratios or
for slopes pooled over several contracts.

In the pricing examples a steeper slope and a smaller error at a fixed budget
are distinct, and they can rank rules differently. The geometric basket is a
ridge function, and low effective dimension alone does not explain the rates of
the discontinuous Asian payoffs. For independent assets, the choice of basis
inside a degenerate PCA eigenspace can change the measured rate, with different
choices favoured by different payoffs, so a covariance matrix alone does not
specify the construction.

Preintegration lowers the exponent of the digital payoff by about $0.5$ in
every fit window. Its main effect on the barrier payoff is on the
size of the error, which is six to nine times smaller at equal sample size,
while the exponent changes by less than four hundredths in every window.
On the arithmetic basket the gain in rate grows with the sample size, so a
single fitted exponent understates the difference at large budgets.

For reporting an empirical rate, state the number of randomisations, choose
the fit window before fitting, and show sensitivity to sub-windows. Attach an
interval that resamples whole randomisations, and report sample kurtosis as a
warning diagnostic. Compare rules through an interval on the slope difference
that preserves any pairing, alongside error ratios at equal sample size and
estimated equal cost. State the cost model and identify any pooling across
contracts. For a deterministic rule, report the slope as descriptive.

\section*{Acknowledgements}

The author thanks the RCP team at EPFL for access to the cluster.

\bibliographystyle{plainnat}
\bibliography{qmcbench_v2}

\appendix

\section{Notation}
\label{app:notation}

\begin{center}
\small
\begin{tabular}{ll@{\qquad}ll}
\toprule
$n$ & sample size & $\alpha,\ \hat\alpha$ & the integral and its estimate \\
$d$ & assets, or a generic dimension & $\beta$ & fitted convergence exponent \\
$m$ & monitoring dates of an Asian contract & $\chi_i$ & exact value of contract $i$ \\
$R$ & independent randomisations & $B$ & barrier level, $130$ here \\
$C$ & contracts in a family & $b$ & base of a sequence \\
$N_B$ & bootstrap draws & $d\,m$ & dimension of a $d$-asset Asian \\
\midrule
$\Bar{S}_{\mathrm{geo}}$ & geometric average at $T$ & $\psi_b$ & radical inverse in base $b$ \\
$\Bar{S}_{\mathrm{ari}}$ & arithmetic average over dates & $\mathcal{D}^*$ & star discrepancy \\
$S^{\mathrm{cl}}_u$ & closed Sobol' index of the set $u$ & $\nu_j$ & direction numbers \\
$S_j$ & first-order index of coordinate $j$ & $\sigma_u^2$ & ANOVA variance component \\
$d_T(\gamma)$ & truncation dimension at level $\gamma$ & $f_u$ & ANOVA term on the set $u$ \\
$f_p$ & calibration integrand with exponent $-p$ & $\widehat{\mathrm{se}}_J$ & jackknife standard error \\
$\beta_{\mathcal{W}}$ & slope of the true RMSE over a window & $\kappa_\ell$ & kurtosis of the error at $n_\ell$ \\
\bottomrule
\end{tabular}
\end{center}

Standard financial symbols are used without redefinition: $K$ strike, $T$
maturity, $r$ risk-free rate, $\sigma_i$ volatility, $\rho$ correlation
between assets, and $\Phi$ the standard normal distribution function.

\section{Proof of Proposition~\ref{prop:slope}}
\label{app:asymptotics}

\begin{proof}
Write $e_\ell^{(r)}$ for the error of randomisation $r$ at $n_\ell$, and let
$m_\ell=R^{-1}\sum_{r=1}^{R}(e_\ell^{(r)})^2$ and
$\bar e_\ell=R^{-1}\sum_{r=1}^{R}e_\ell^{(r)}$. The vectors
$\bigl((e_1^{(r)})^2,\dots,(e_L^{(r)})^2\bigr)$ are independent and identically
distributed over $r$, with mean $\sigma^2=(\sigma_1^2,\dots,\sigma_L^2)$ and,
because the fourth moments are finite, with a finite covariance matrix $\Sigma$,
$\Sigma_{\ell k}=\operatorname{Cov}(e_\ell^2,e_k^2)$. By the multivariate
central limit theorem, $\sqrt{R}\,(m-\sigma^2)$ converges in distribution to
$\mathcal{N}(0,\Sigma)$.

Each randomisation is unbiased, so $\mathbb{E}\,e_\ell=0$, and
\[
  s_\ell^2=\frac{1}{R-1}\sum_{r=1}^{R}\bigl(e_\ell^{(r)}-\bar e_\ell\bigr)^2
  =\frac{R}{R-1}\bigl(m_\ell-\bar e_\ell^{\,2}\bigr).
\]
Since $\sqrt{R}\,\bar e_\ell$ is bounded in probability,
$\sqrt{R}\,\bar e_\ell^{\,2}$ converges to zero in probability, and so does
$\sqrt{R}\,(s_\ell^2-m_\ell)$. Hence $\sqrt{R}\,(s^2-\sigma^2)$ also converges
in distribution to $\mathcal{N}(0,\Sigma)$.

Let $g(u)=\tfrac12\sum_{\ell}w_\ell\log u_\ell$ for $u$ with positive
components. Then $\hat\beta=g(s^2)$ and $\beta_{\mathcal{W}}=g(\sigma^2)$, and
$g$ is differentiable at $\sigma^2$ with partial derivatives
$w_\ell/(2\sigma_\ell^2)$. By the delta method,
$\sqrt{R}\,(\hat\beta-\beta_{\mathcal{W}})$ converges in distribution to a
normal law with mean zero and variance
$\nabla g(\sigma^2)^{\top}\,\Sigma\,\nabla g(\sigma^2)=v_{\mathcal{W}}$. On the
diagonal,
$\Sigma_{\ell\ell}=\mathbb{E}\,e_\ell^4-\sigma_\ell^4=\sigma_\ell^4(\kappa_\ell-1)$.

For a family of $C$ contracts that share each randomisation, the pooled
variance is
\[
  s_\ell^2=\frac{R}{R-1}\,\frac{1}{C}\sum_{c=1}^{C}
  \bigl(m_{\ell,c}-\bar e_{\ell,c}^{\,2}\bigr),
\]
and the same argument applies to the vectors whose components are
$C^{-1}\sum_{c}(e_{\ell,c}^{(r)})^2$.

For two rules that share each randomisation, apply the argument to the vectors
of length $2L$ that hold the squared errors of both rules. They are independent
and identically distributed over $r$, whatever the dependence between the two
rules within one randomisation. The difference of the slopes is
$g(s_B^2)-g(s_A^2)$, with gradient
$\bigl(-\nabla g(\sigma_A^2),\nabla g(\sigma_B^2)\bigr)$ at
$(\sigma_A^2,\sigma_B^2)$, and the quadratic form of this gradient in the
covariance matrix of the stacked vectors equals $v_A+v_B-2c_{AB}$. If the rules
are randomised independently, the covariances between their squared errors
vanish and $c_{AB}=0$.
\end{proof}

\section{Closed form for the geometric basket}
\label{app:closedform}

Let every asset start at $S(0)$. By Section~\ref{sec:n_simu},
$\Bar{S}_{\mathrm{geo}}=S(0)\exp\{(r-q-\tfrac12\bar\sigma^2)T+\bar\sigma\sqrt{T}\,Z\}$
with $Z$ standard normal, where
\[
  \bar\sigma=\frac{1}{d}\Bigl(\sum_{i=1}^d\sum_{j=1}^d\rho_{ij}\sigma_i\sigma_j\Bigr)^{1/2},
  \qquad
  q=\frac{1}{2d}\sum_{i=1}^d\sigma_i^2-\frac{\bar\sigma^2}{2}.
\]
The geometric average is therefore distributed like a lognormal asset with
volatility $\bar\sigma$ and dividend yield $q$, and the Black--Scholes formula
gives
\[
  \mathbb{E}\bigl[e^{-rT}(\Bar{S}_{\mathrm{geo}}-K)^+\bigr]
  =S(0)\,e^{-qT}\,\Phi(\delta)-K\,e^{-rT}\,\Phi(\delta-\bar\sigma\sqrt{T}),
\]
where
\[
  \delta=\frac{\log(S(0)/K)+(r-q+\tfrac12\bar\sigma^2)T}{\bar\sigma\sqrt{T}}.
\]
This expression assumes $\bar\sigma\sqrt{T}>0$. If it is zero, the price
is the discounted positive part of the deterministic geometric average
minus the strike.

\section{Faure errors and sequence base}
\label{app:faure}

The base $b$ of a Faure sequence grows with the dimension
(Section~\ref{sec:faure-base}) and its equidistribution holds for blocks of
$b^k$ points, so at a fixed sample size a larger base means fewer complete
blocks. With $n=5000$, the sample contains a full block of $b^2$ points only
while $d$ is at most $67$. The integrand is a geometric basket call in
dimension $d$ with independent assets, $S_i(0)=100$, $r=5\%$, $T=1$, volatility
$31\%$ and $K=100$. The error of the deterministic Faure sequence, which starts
at its first point, is its absolute error against the closed-form price, and
the errors of the randomised rules are root mean square errors against that
price over $512$ randomisations. The error of the deterministic sequence rises
once $n/b^2$ falls to a few units, and scrambling removes most of that rise
(Table~\ref{tab:faure}). This points to the base relative to
the sample size as an explanation of the loss. The sweep changes both the
integrand and the base, so it does not isolate the base effect experimentally.

\begin{table}[H]
\centering
\begin{tabular}{rrrrccc}
\toprule
$d$ & base $b$ & $b^2$ & $n/b^2$ & Faure & Faure & Sobol' \\
 & & & & (deterministic) & (scrambled) & (scrambled) \\
\midrule
8 & 11 & 121 & 41.3 & 0.0312 & 0.0141 & 0.0105 \\
16 & 17 & 289 & 17.3 & 0.0373 & 0.0118 & 0.0116 \\
32 & 37 & 1369 & 3.65 & 0.2855 & 0.0102 & 0.0095 \\
64 & 67 & 4489 & 1.11 & 0.7935 & 0.0073 & 0.0066 \\
96 & 97 & 9409 & 0.531 & 0.9846 & 0.0093 & 0.0060 \\
128 & 131 & 17161 & 0.291 & 0.7346 & 0.0102 & 0.0052 \\
160 & 163 & 26569 & 0.188 & 0.5353 & 0.0100 & 0.0050 \\
256 & 257 & 66049 & 0.0757 & 0.1054 & 0.0076 & 0.0044 \\
512 & 521 & 271441 & 0.0184 & 1.0304 & 0.0056 & 0.0036 \\
1024 & 1031 & 1062961 & 0.0047 & 2.6115 & 0.0041 & 0.0030 \\
\bottomrule
\end{tabular}

\caption{Error against the closed-form price of a geometric basket call, as a
function of the dimension at fixed $n=5000$. The deterministic Faure
sequence has error $0.0312$ at $d=8$, rises to $0.9846$ at $d=96$ and is erratic
beyond, for instance $0.1054$ at $d=256$. The scrambled sequence stays within a
factor of four of its smallest error at every dimension.
Scrambled Sobol' points are shown for comparison.}
\label{tab:faure}
\end{table}

\begin{figure}[H]
\centering
\includegraphics[width=0.82\textwidth]{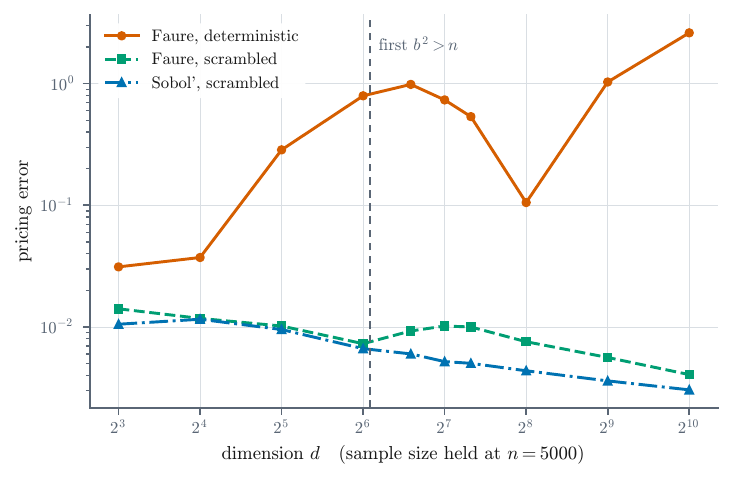}
\caption{Error against dimension at $n=5000$. The dashed line marks $d=68$, the first
dimension for which $b^2>5000$. The deterministic error is erratic and grows
substantially at several larger dimensions. Randomisation avoids most of this
increase.}
\label{fig:faure}
\Description{Pricing errors against dimension at fixed sample size. The
deterministic Faure curve is erratic, while scrambling substantially reduces
its error. A vertical marker identifies the first dimension at which
the square of the Faure base exceeds the sample size.}
\end{figure}

\section{Mean and median aggregation}
\label{app:median}

Pan and Owen~\citep{PanOwen2022} show that the median of independent randomised
net estimates can converge much faster than their mean for sufficiently smooth
integrands. We tested whether the median helps on the digital and barrier
payoffs of the central Asian contract under principal components, which are not
smooth and so lie outside that result. Each estimate
combines nine independent randomisations of the same $n$-point net, either by
their mean or by their median, so the two estimators use exactly the same
function values. At these sample sizes the two show no clear difference
(Table~\ref{tab:median}). At every discretisation their intervals overlap, and
the median has the better exponent on four of the six rows. Overlapping
intervals do not show that the estimators are equivalent, and we did not test
equivalence against a stated margin.

\begin{table}[H]
\centering
\begin{tabular}{lccc}
\toprule
Payoff & $m$ & Mean of 9 groups & Median of 9 groups \\
\midrule
Digital & 64 & $-0.589$ {\scriptsize $[-0.627,\,-0.549]$} & $-0.609$ {\scriptsize $[-0.643,\,-0.575]$} \\
Barrier & 64 & $-0.563$ {\scriptsize $[-0.598,\,-0.528]$} & $-0.557$ {\scriptsize $[-0.589,\,-0.525]$} \\
Digital & 121 & $-0.580$ {\scriptsize $[-0.611,\,-0.542]$} & $-0.599$ {\scriptsize $[-0.633,\,-0.564]$} \\
Barrier & 121 & $-0.521$ {\scriptsize $[-0.559,\,-0.488]$} & $-0.542$ {\scriptsize $[-0.574,\,-0.511]$} \\
Digital & 256 & $-0.562$ {\scriptsize $[-0.598,\,-0.526]$} & $-0.569$ {\scriptsize $[-0.603,\,-0.536]$} \\
Barrier & 256 & $-0.577$ {\scriptsize $[-0.617,\,-0.535]$} & $-0.564$ {\scriptsize $[-0.598,\,-0.528]$} \\
\bottomrule
\end{tabular}

\caption{Convergence exponent of the mean and of the median of nine
independent randomisations, from $256$ repetitions at $n=2^{10},\dots,2^{15}$
points per randomisation. The error is measured against an independent
reference value computed with preintegration from $64$ randomisations of
$2^{20}$ points, whose standard error is below one per cent of the error of the
mean at the largest sample size. Intervals are BCa and condition on this
reference. Its uncertainty is not propagated into them.}
\label{tab:median}
\end{table}

\end{document}